\documentclass[12pt,a4paper]{article}
\usepackage{amssymb}
\usepackage{amsfonts}
\usepackage{graphicx}
\usepackage{graphics}
\usepackage{epsfig}
\usepackage{rotating}
\usepackage{amsmath}
\usepackage{url}
\usepackage{multicol}
\usepackage{float}
\usepackage{empheq}
\usepackage{latexsym}
\usepackage{cite}
\usepackage{physics}
\begin{document}

\title{\begin{flushright}
{\small INR-TH-2016-043}
\end{flushright} {\bf Problem with classical stability\\ of U(1) gauged Q-balls}}

\author{
 A.G.~Panin$^{a,b}$, M.N.~Smolyakov$^{c,a}$
\\
$^a${\small{\em
Institute for Nuclear Research of the Russian Academy
of Sciences,}}\\
{\small{\em 60th October Anniversary prospect 7a, Moscow 117312,
Russia
}}\\
$^b${\small{\em Moscow Institute of Physics and Technology,}}\\
{\small{\em Institutsky per. 9, Dolgoprudny 141700, Russia}}\\
$^c${\small{\em Skobeltsyn Institute of Nuclear Physics, Lomonosov Moscow
State University,
}}\\
{\small{\em Moscow 119991, Russia}}}

\date{}
\maketitle

\begin{abstract}
In this paper, we present a detailed study of the problem of classical stability of U(1) gauged Q-balls. In particular, we show that the standard methods that are suitable for establishing the classical stability criterion for ordinary (nongauged) one-field and two-field Q-balls are not effective in the case of U(1) gauged Q-balls, although all the technical steps of calculations can be performed in the same way as those for ordinary Q-balls. We also present the results of numerical simulations in models with different scalar field potentials, explicitly demonstrating that, in general, the regions of stability of U(1) gauged Q-balls are not defined in the same way as in the case of ordinary Q-balls. Consequently, the classical stability criterion for ordinary Q-balls cannot be applied to U(1) gauged Q-balls in the general case.
\end{abstract}

\section{Introduction}
A class of nontopological solitons, initially proposed in \cite{Rosen0} and known as Q-balls \cite{Coleman:1985ki}, has been widely discussed in the literature during the past years. It is interesting to mention that generalization from the global $U(1)$ symmetry to the gauge $U(1)$ symmetry was proposed in the paper \cite{Rosengauged}, which was published together with \cite{Rosen0} (in the same issue of the journal). Later such a theory was examined in the well-known paper \cite{Lee:1988ag}. One can also recall papers \cite{BF,BF1,Gulamov:2013cra}, where gauged Q-balls\footnote{For simplicity, from here on, we call $U(1)$ gauged Q-balls ``gauged Q-balls''.} were examined mainly from a theoretical point of view, as well as papers \cite{Lee:1991bn,Arodz:2008nm,Dzhunushaliev:2012zb,Tamaki:2014oha,Brihaye:2014gua,Hong:2015wga,Gulamov:2015fya}, where solutions for gauged Q-ball were obtained numerically.

Stability of such nontopological soliton solutions, in particular, the classical stability, is an important problem. The well-known classical stability criterion for ordinary (nongauged) Q-balls, which is $\frac{dQ}{d\omega}<0$, where $Q$ is the Q-ball charge, was derived in \cite{LeePang,Friedberg:1976me}. One may suppose that analogous classical stability criterion is valid for $U(1)$ gauged Q-balls. However, there is no rigorous mathematical proof supporting this hypothesis.

In the present paper we will try to derive the classical stability criterion for such gauged Q-balls. Instead of using the approach of \cite{LeePang,Friedberg:1976me}, for the calculations we will apply the Vakhitov-Kolokolov method \cite{VK,Kolokolov}, which was used for derivation of the classical stability criterion for the systems described by the nonlinear Schr\"{o}dinger equation (for a more detailed discussion of the stability of localized solutions in the systems described by the nonlinear Schr\"{o}dinger equation, see \cite{Makhankov:1978rg,Makhankov,AA}). We will show that this method works very well for the case of ordinary one-field and two-field Q-balls, but does not provide any information about the possible classical stability criterion of $U(1)$ gauged Q-balls, although all the technical steps of calculations can be performed in this case too. We will also provide numerical simulations of classical (in)stabilities in models with different scalar field potentials, supporting the results of analytical considerations.

The paper is organized as follows. As a demonstration of the  Vakhitov-Kolokolov method, in Section~2 we present the proof of the classical stability criterion for the case of ordinary one-field and two-field Q-balls. In Section~3 we make an attempt to derive the classical stability criterion for the case of $U(1)$ gauged Q-balls and discuss the reasons why this attempt is unsuccessful. We also present the results of numerical simulations of classical (in)instabilities of gauged Q-balls. In the Conclusion we briefly discuss the obtained results.

\section{Classical stability of ordinary Q-balls}
In this section, we will derive the classical stability criterion for the case of one-field and two-field Q-balls.\footnote{Although solutions in two-field models like the one of \cite{Friedberg:1976me} are not Q-balls in the sense of Coleman's definition of Q-balls \cite{Coleman:1985ki}, they are of the same kind, so we also call such soliton solutions ``Q-balls''.} For the first time, this criterion was derived in \cite{LeePang} for one-field Q-balls and in \cite{Friedberg:1976me} for the model with two scalar fields, which was proposed and examined in that paper. The proof of \cite{LeePang,Friedberg:1976me} was based on examining the properties of the energy functional of the system, while keeping the charge fixed. As was noted in the Introduction, instead of using the same approach, we will apply the Vakhitov-Kolokolov method \cite{VK,Kolokolov}, which was used to obtain the classical stability criterion for the systems described by the nonlinear Schr\"{o}dinger equation. This method is based on the use of only the linearized equations of motion for the perturbations above the background solution. We will apply this method to the case of $U(1)$ gauged Q-balls, but first we will demonstrate how it works in the simpler cases of ordinary one-field and two-field Q-balls (from the mathematical point of view, two-field Q-balls are closer to the gauged Q-balls than the one-field Q-balls, so the explicit examination of this case makes sense). So, let us proceed to the case of one-field Q-balls, which we will study in detail.

\subsection{One-field Q-balls}
Let us start with the standard action, describing a complex scalar field $\phi$ in the flat (d+1)-dimensional space-time with the coordinates $x^{\mu}=\{t,\vec x\}$, $\mu=0,1,...,d$; in the form
\begin{equation}
S=\int dtd^{d}x\left(\partial_\mu\phi^*\partial^\mu\phi-V(\phi^*\phi)\right).
\label{sys}
\end{equation}
For a Q-ball solution, we consider the standard ansatz
\begin{equation}\label{qballsolution}
\phi(t,\vec x)=e^{i\omega t}f(r),
\end{equation}
where $r=\sqrt{{\vec x}^{2}}$ and $f(r)$ is a real function which is supposed to have no nodes (without loss of generality, we can set $f(r)>0$ for any $r$) and to satisfy the conditions
\begin{equation}
\partial_{r}f(r)|_{r=0}=0,\qquad \lim\limits_{r\to\infty}f(r)=0.
\end{equation}
In this case, the function $f(r)$ satisfies the equation
\begin{equation}\label{eqqball}
\omega^{2}f+\Delta f-\frac{dV}{d(\phi^{*}\phi)}\biggl|_{\phi^{*}\phi=f^{2}}f=0,
\end{equation}
where $\Delta=\sum\limits_{i=1}^{d}\partial_{i}\partial_{i}$. The charge of a Q-ball is defined in the standard way as
\begin{equation}\label{qballcharge}
Q=i\int\left(\phi\partial_{0}\phi^*-\phi^*\partial_{0}\phi\right)d^{d}x=2\omega\int f^{2}d^{d}x.
\end{equation}

In order to examine the classical stability of a Q-ball, on should consider small perturbations above the background solution \eqref{qballsolution} and examine the corresponding linearized equations of motion. The standard ansatz for the perturbations takes the form \cite{Anderson:1970et,MarcVent}
\begin{equation}
\phi(t,\vec x)=e^{i\omega t}f(r)+e^{i\omega t}\left(a(\vec x)e^{i\rho t}+b(\vec x)e^{-i\rho^{*} t}\right).
\end{equation}
Note that the functions $a(\vec x)$ and $b(\vec x)$ are not supposed to be spherically symmetric. From the very beginning, it is convenient to use the notations
\begin{equation}
\xi_{1}(\vec x)=a(\vec x)+b^{*}(\vec x),\qquad \xi_{2}(\vec x)=a(\vec x)-b^{*}(\vec x).
\end{equation}
With these notations, the corresponding linearized equations of motion take the form
\begin{eqnarray}\label{lineq1}
L_{1}\xi_{1}-2\omega\rho\xi_{2}-\rho^{2}\xi_{1}=0,\\ \label{lineq2}
L_{2}\xi_{2}-2\omega\rho\xi_{1}-\rho^{2}\xi_{2}=0,
\end{eqnarray}
where the operators $L_{1}$ and $L_{2}$ are defined as
\begin{eqnarray}
&&L_{1}=-\Delta+U(r)+2S(r)-\omega^{2},\\ \label{L2eq}
&&L_{2}=-\Delta+U(r)-\omega^{2}
\end{eqnarray}
with
\begin{equation}\label{UGdef}
U(r)=\frac{dV}{d(\phi^{*}\phi)}\biggl|_{\phi^{*}\phi=f^{2}(r)},\qquad S(r)=\frac{d^{2}V}{d(\phi^{*}\phi)^{2}}\biggl|_{\phi^{*}\phi=f^{2}(r)}f^{2}(r).
\end{equation}
Possible instabilities are described by solutions to equations \eqref{lineq1}, \eqref{lineq2} with $\textrm{Im}\,\rho\neq 0$.

Using equations \eqref{lineq1} and \eqref{lineq2} it is possible to show (see the detailed derivation in Appendix~A) that the following relation fulfills
\begin{equation}
\rho^{2}={\rho^{*}}^2.
\end{equation}
This equation has two obvious solutions: $\rho=\gamma$ and $\rho=i\gamma$, where $\gamma$ is a real constant. Since we are interested in classical instabilities, below we will consider only the case $\rho=i\gamma$.

As was noted above, the unstable modes can only have the form
\begin{equation}
\phi(t,\vec x)=e^{i\omega t}f(r)+e^{i\omega t}e^{\gamma t}\left(u(\vec x)+iv(\vec x)\right),
\end{equation}
where $\gamma$ is a real constant and $u(\vec x)$, $v(\vec x)$ are real functions. In what follows, we will consider $\gamma\neq 0$ and use the notations $u$, $v$ instead of $\xi_{1}$ and $\xi_{2}$. In these notations, the linearized equations of motion take the form
\begin{eqnarray}
\Delta u+\omega^{2}u+2\omega\gamma v-\gamma^{2}u-Uu-2Su=0,\\ \label{equv2}
\Delta v+\omega^{2}v-2\omega\gamma u-\gamma^{2}v-Uv=0.
\end{eqnarray}
Let us introduce the operators $L_{+}$ and $L_{-}$ defined by
\begin{eqnarray}
&&L_{+}=L_{1}=-\Delta+U(r)+2S(r)-\omega^{2},\\
&&L_{-}=L_{2}+\gamma^{2}=-\Delta+U(r)-\omega^{2}+\gamma^{2},
\end{eqnarray}
and the operator
\begin{equation}
\hat L=\begin{pmatrix}
L_{+} & 0 \\
0 & L_{-}
\end{pmatrix}.
\end{equation}
Note that we have added $\gamma^{2}$ to the definition of the operator $L_{-}$. The necessity for this step will become clear later.

In the operator form, the linearized equations of motion for the functions $u$ and $v$ look like
\begin{equation}\label{eqmatrixform}
\begin{pmatrix}
L_{+} & 0 \\
0 & L_{-}
\end{pmatrix}\Psi=-\gamma^{2}\begin{pmatrix}u \\ 0\end{pmatrix}+2\omega\gamma\begin{pmatrix}v \\ -u\end{pmatrix},
\end{equation}
where the notation
\begin{equation}
\Psi=\begin{pmatrix}u \\ v\end{pmatrix}
\end{equation}
is introduced for convenience.

Now let us look at the properties of perturbations, following from the linearized equations of motion. First, by multiplying equation \eqref{equv2} by $f$, integrating the result over the spatial volume and using the fact that $L_{2}f=0$ (see equation \eqref{eqqball}), we get
\begin{equation}\label{extracond1}
\bra{\Phi}\ket{\Psi}=\int \Phi^{T}\Psi d^{d}x=0,
\end{equation}
where
\begin{equation}
\Phi=\begin{pmatrix}2\omega f \\ \gamma f\end{pmatrix}.
\end{equation}
Relation \eqref{extracond1} is simply the consequence of the total charge conservation.

Second, there are obvious zero modes (i.e., the eigenstates with the zero eigenvalue, not the lowest modes) of the operator $\hat L$:
\begin{equation}\label{zeromode}
\hat L T_{i}=\hat L\begin{pmatrix}\partial_{i}f \\ 0\end{pmatrix}=0.
\end{equation}
The existence of the eigenstates $T_{i}$, $i=1,...,d$, is the corollary of the fact that $L_{+}\partial_{i}f=0$, whereas $\partial_{i}f$ are just the translational modes. Note that since $f(r)>0$ for any $r$, it is the eigenfunction of the lowest eigenstate of the operator $L_{-}$ with the eigenvalue $\gamma^{2}>0$ (for example, for $d=3$ it corresponds to the $1s$ level in the spherically symmetric quantum mechanical potential $U(r)-\omega^{2}+\gamma^{2}$), so all the eigenstates of the operator $\hat L$ with the zero and negative eigenvalues are defined by the corresponding eigenstates of the operator $L_{+}$.\footnote{This is the first reason to add $\gamma^{2}$ to the definition of the operator $L_{-}$ --- all eigenvalues of $L_{-}$ are positive and there are no zero modes of the operator $\hat L$ coming from $L_{-}$.} At this point we should impose the following restriction on the spectrum of the operator $L_{+}$, which is crucial for the subsequent proof:
\begin{itemize}
\item
the operator $L_{+}$ has {\em only one} negative eigenvalue.
\end{itemize}
For the one-dimensional case $d=1$ this restriction fulfills automatically. Indeed, the translational mode $\partial_{x}f$ has only one node, thus there is only one eigenstate with negative eigenvalue \cite{LandauLifshitz}. However, this is not so for the case $d>1$. For example, for $d=3$ the translational modes $\partial_{i}f$ form the $2p$ level and it is possible that this level is energetically higher than, say, $1s$ and $2s$ levels (for an example of such situation in atomic physics, see \cite{Connerade}). However, the restriction on the number of eigenstates of $L_{+}$ with negative eigenvalues seems to be fulfilled at least for the simple scalar field potentials which are usually considered for Q-balls.

And third, from equation \eqref{eqmatrixform} it follows that
\begin{equation}
\expval{\hat L}{\Psi}=-\gamma^{2}\bra{u}\ket{u}.
\end{equation}
It is clear that if $\expval{\hat L}{\Psi}>0$, then there are no unstable modes above the Q-ball solution. So, the question is under which conditions the lowest possible value of $\expval{\hat L}{\Psi}$ is positive, taking into account the extra condition \eqref{extracond1}? This question leads to the variational problem for the functional
\begin{equation}
I=\expval{\hat L}{\Psi}-\theta\bra{\Phi}\ket{\Psi}-\lambda\left(\bra{\Psi}\ket{\Psi}-C^{2}\right),
\end{equation}
where $\theta$ and $\lambda$ are the Lagrange multipliers, the second Lagrange multiplier $\lambda$ stands for the normalization condition $\bra{\Psi}\ket{\Psi}=C^{2}$, where $C\neq 0$ is a real constant of the corresponding dimension. The standard variational procedure leads to the equation
\begin{equation}\label{vareq}
2\hat L\Psi-2\lambda\Psi-\theta\Phi=0.
\end{equation}
Now we take the orthonormal system of eigenfunctions of the operator $\hat L$, which is supposed to be full, and decompose $\Psi$ and $\Phi$ in these eigenfunctions. Let us label the lowest eigenfunction and eigenvalue of the operator $\hat L$ as $\psi_{-1}$ and $\lambda_{-1}$, respectively, where $\lambda_{-1}<0$ (as was noted above, we suppose that there is only one such eigenvalue), the zero modes $\psi_{0,i}\sim T_{i}$ correspond to the eigenvalue $\lambda_{0}=0$, and all the other modes $\psi_{n}$ with $n>0$ are such that $\lambda_{n}>0$. We also choose $\psi_{n}$ such that they are dimensionless and $\bra{\psi_{n}}\ket{\psi_{n}}=1$ for any $n$.

It is clear that $\bra{T_{i}}\ket{\Phi}=0$, so the decomposition of $\Phi$ takes the form
\begin{equation}\label{Phidec}
\Phi=\sum\limits_{\substack{n=-1\\n\neq 0}}^{\infty}d_{n}\psi_{n},
\end{equation}
where $d_{n}$ are the coefficients of the decomposition.\footnote{Of course, the set of eigenstates of the operator $\hat L$ is not necessarily discrete; there may exist a continuous part of the spectrum. So, the most general form of the decomposition looks like $\Phi(\vec x)=\sum\limits_{\substack{n=-1\\n\neq 0}}^{N}d_{n}\psi_{n}(\vec x)+\int\limits_{\lambda_{c}}^{\infty}d(\tilde\lambda)\psi(\tilde\lambda,\vec x)d\tilde\lambda$. However, the existence of the continuous part of the spectrum does not change the proof if $\lambda_{c}>\lambda_{1}>0$. So, for the sake of simplicity, below we will use the decomposition \eqref{Phidec} instead of the more general one.} Thus, from equation \eqref{vareq} we get
\begin{equation}\label{Psidec}
\Psi=\frac{\theta}{2}\sum\limits_{\substack{n=-1\\n\neq 0}}^{\infty}\frac{d_{n}}{\lambda_{n}-\lambda}\psi_{n}.
\end{equation}
Multiplying equation \eqref{vareq} by $\Psi^{T}$, integrating the result over the spatial volume and using the conditions \eqref{extracond1} and $\bra{\Psi}\ket{\Psi}=C^{2}$, we obtain
\begin{equation}\label{positivity}
\expval{\hat L}{\Psi}=\lambda\bra{\Psi}\ket{\Psi}=\lambda C^{2}.
\end{equation}
Thus, the sign of $\expval{\hat L}{\Psi}$ is defined by the sign of $\lambda$.

\begin{figure}
\begin{tabular}{cc}
\includegraphics[width=0.47\linewidth]{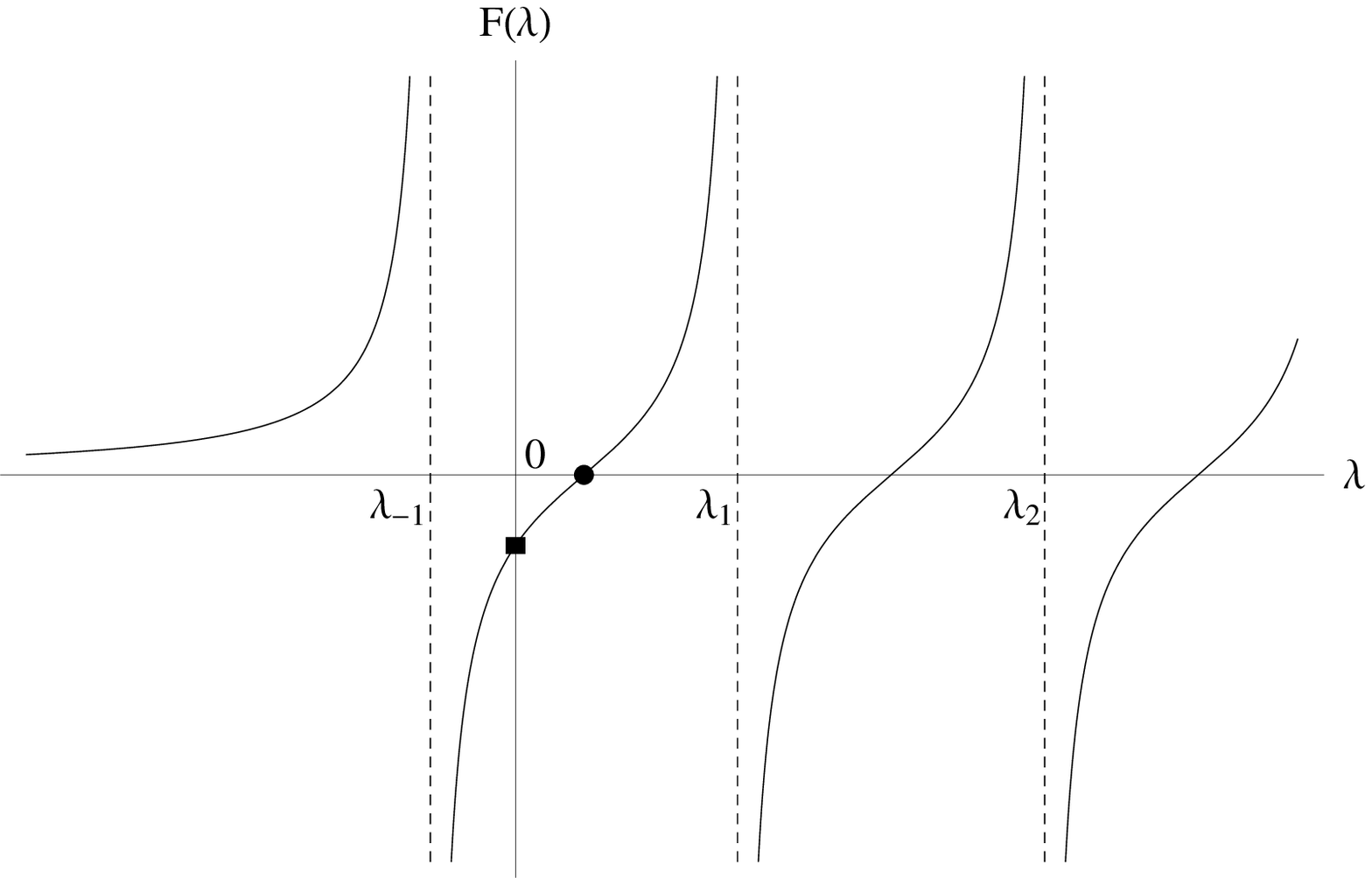}&
\includegraphics[width=0.47\linewidth]{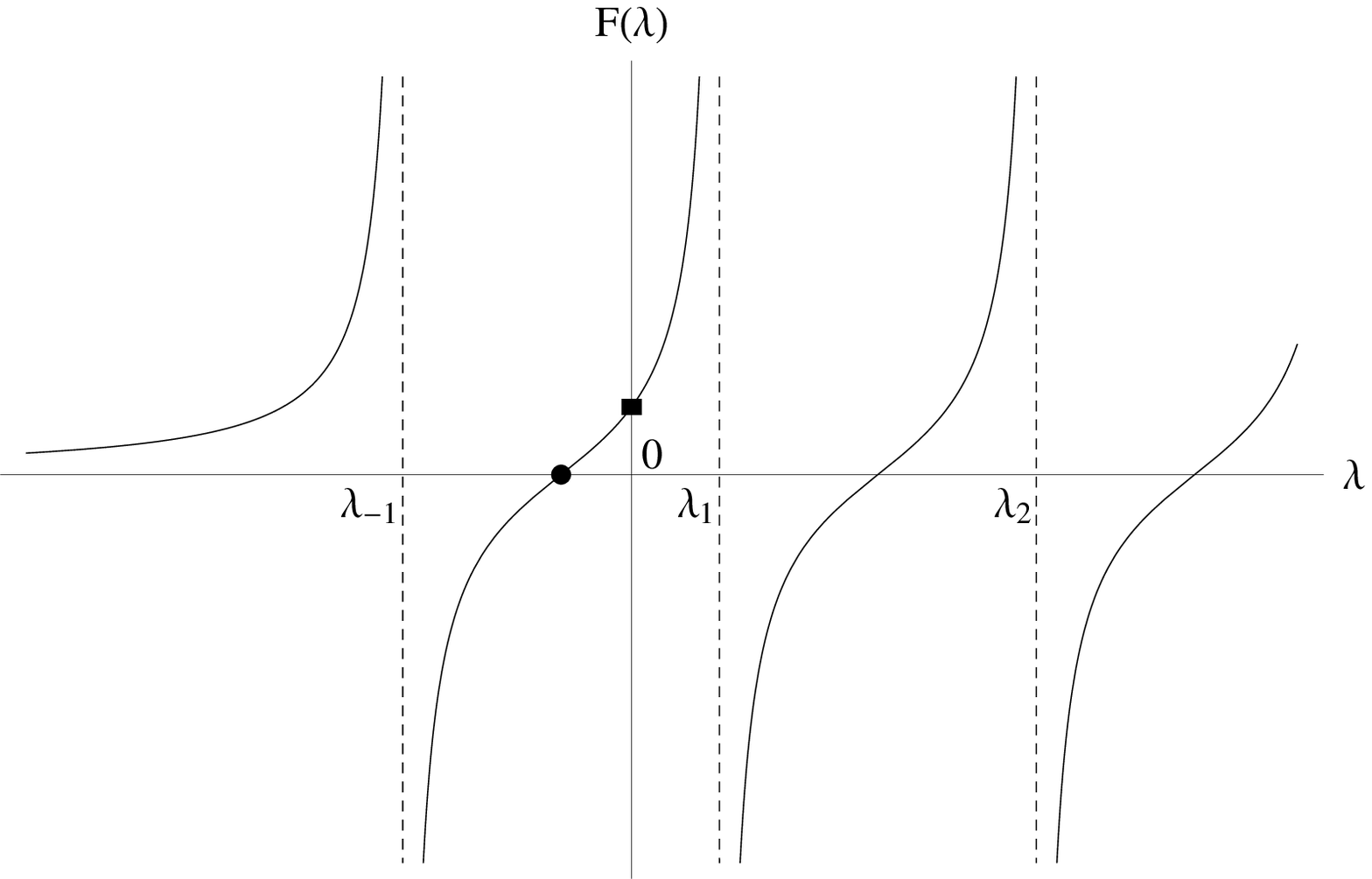}\\
$(a)$&$(b)$\\
\end{tabular}
\caption{Schematic representation of the function $F(\lambda)$ for the cases with $F(0)<0$ ($a$) and $F(0)>0$ ($b$). The dashed lines correspond to the asymptotes $\lambda=\lambda_{n}$, the dots mark the smallest solutions of equation $F(\lambda)=0$, and the squares mark the points $(0,F(0))$.}
\label{figure1}
\end{figure}

The next step of the proof is to find a minimal value of $\lambda$. To this end, we take equation \eqref{extracond1} and substitute \eqref{Phidec} and \eqref{Psidec} into it. We get the equation
\begin{equation}
\frac{\theta}{2}\sum\limits_{\substack{n=-1\\n\neq 0}}^{\infty}\frac{d_{n}^{2}}{\lambda_{n}-\lambda}=0,
\end{equation}
which defines possible values of $\lambda$ for which the extra condition \eqref{extracond1} is satisfied. Let us define the function
\begin{equation}
F(\lambda)=\sum\limits_{\substack{n=-1\\n\neq 0}}^{\infty}\frac{d_{n}^{2}}{\lambda_{n}-\lambda}
\end{equation}
It is clear that everywhere except the points $\lambda_{n}$, the function $F(\lambda)$ is a monotonically growing function. A schematic representation of this function is presented in Fig.~\ref{figure1}.
It is clear that, due to the monotonic growth of $F(\lambda)$, the minimal possible value of the solution to equation $F(\lambda)=0$ is positive for $F(0)<0$; see Fig.~\ref{figure1}. So, if $F(0)<0$, then there cannot be unstable modes in the perturbations above the Q-ball solution.

The last step is to represent the value of $F(0)$ through the known characteristics of the Q-ball. In order to do it, we write $F(0)$ as
\begin{equation}\label{F0def}
F(0)=\sum\limits_{\substack{n=-1\\n\neq 0}}^{\infty}\frac{d_{n}^{2}}{\lambda_{n}}=\expval{{\hat L}^{-1}}{\Phi}.
\end{equation}
Next, we observe that by differentiating equation \eqref{eqqball} with respect to $\omega$, we get
\begin{equation}
L_{+}\frac{df}{d\omega}=2\omega f.
\end{equation}
The latter allows to construct the column
$\begin{pmatrix}\frac{df}{d\omega} \\ \frac{f}{\gamma}\end{pmatrix}$,
for which the relation
\begin{equation}\label{LPhieq}
\hat L \begin{pmatrix}\frac{df}{d\omega} \\ \frac{f}{\gamma}\end{pmatrix}=\begin{pmatrix}2\omega f \\ \gamma f\end{pmatrix}=\Phi.
\end{equation}
fulfills,\footnote{This is the second reason to add $\gamma^{2}$ to the definition of the operator $L_{-}$.} leading to
\begin{equation}
\begin{pmatrix}\frac{df}{d\omega} \\ \frac{f}{\gamma}\end{pmatrix}={\hat L}^{-1}\Phi.
\end{equation}
Substituting the latter formula into \eqref{F0def}, we arrive at
\begin{equation}
F(0)=\int \begin{pmatrix}2\omega f & \gamma f\end{pmatrix}\begin{pmatrix}\frac{df}{d\omega} \\ \frac{f}{\gamma}\end{pmatrix}d^{d}x=
\int \left(2\omega f\frac{df}{d\omega}+f^{2}\right)d^{d}x=\frac{1}{2}\frac{dQ}{d\omega},
\end{equation}
where we have used the definition of the Q-ball charge \eqref{qballcharge}. Thus, if $\frac{dQ}{d\omega}<0$, the minimal possible value of $\lambda$ such that the condition \eqref{extracond1} holds, is positive, leading to the positivity of the minimal possible value of $\expval{\hat L}{\Psi}$ in accordance with \eqref{positivity}. The latter means that there are no unstable modes with $\gamma\neq 0$ for $\frac{dQ}{d\omega}<0$. This ends the proof.

The following remarks are in order.
\begin{itemize}
\item
The condition $\frac{dQ}{d\omega}>0$ does not mean that there exist unstable modes --- the results presented above and the results of \cite{Friedberg:1976me,LeePang} (analogously to the corresponding results of \cite{VK,Kolokolov} for the nonlinear Schr\"{o}dinger equation) only state that there are no unstable modes if $\frac{dQ}{d\omega}<0$ (of course, if the restriction on the number of negative eigenvalues of the operator $L_{+}$ is fulfilled). However, the explicit analyses of instabilities in different models, which were carried out in \cite{Anderson:1970et,MarcVent,Gulamov:2013ema}, show that at least in the models discussed in these papers there do exist unstable modes for $\frac{dQ}{d\omega}>0$.
\item
What if there are two different negative eigenvalues of the operator $L_{+}$ (and, consequently, of the operator $\hat L$)? This situation is schematically represented in Fig.~\ref{figure2}. One can see that in this case there always exists negative solution to the equation $F(\lambda)=0$, which invalidates the classical stability criterion $\frac{dQ}{d\omega}<0$.
\begin{figure}
\center{\includegraphics[width=0.5\linewidth]{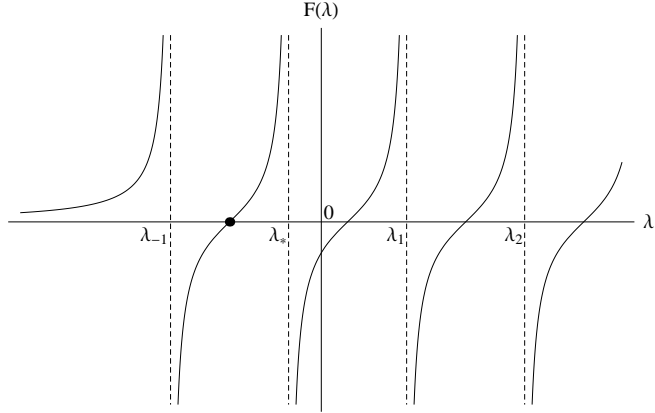}}
\caption{Schematic representation of the function $F(\lambda)$ for the case with the extra negative eigenvalue $\lambda_{*}$ of the operator $\hat L$. The dashed lines correspond to the asymptotes $\lambda=\lambda_{n}$ and $\lambda=\lambda_{*}$; the dot labels the smallest solution of equation $F(\lambda)=0$.}
\label{figure2}
\end{figure}
Meanwhile, we are not aware of any explicit example of a Q-ball, for which the operator $L_{+}$ has more than one negative eigenvalues and which is classically unstable for $\frac{dQ}{d\omega}<0$.
\item
What if there is another zero mode $\psi_{0}^{(+)}$ of the operator $L_{+}$? Then, the operator $\hat L$ has the zero mode $\psi_{0}=\begin{pmatrix}\psi_{0}^{(+)} \\ 0\end{pmatrix}$. Using \eqref{LPhieq}, it is easy to show that $\bra{\psi_{0}}\ket{\Phi}=0$, so this mode does not contribute to the decomposition \eqref{Phidec}.
\item
In principle, it is possible that the positive part of the spectrum of the operator $\hat L$ is purely continuous and starts at $\lambda_{c}>0$. Then, for any $\lambda$ such that $\lambda_{-1}<\lambda<0$ we have
\begin{equation}\label{onlycontspectrum}
F(\lambda)=\frac{d_{-1}^{2}}{\lambda_{-1}-\lambda}+\int\limits_{\lambda_{c}}^{\infty}\frac{d^{2}(\tilde\lambda)}{\tilde\lambda-\lambda}d\tilde\lambda<
\frac{d_{-1}^{2}}{\lambda_{-1}}+\int\limits_{\lambda_{c}}^{\infty}\frac{d^{2}(\tilde\lambda)}{\tilde\lambda}d\tilde\lambda=
F(0)=\frac{1}{2}\frac{dQ}{d\omega},
\end{equation}
where $\lambda_{-1}<0$. Thus, if $\frac{dQ}{d\omega}<0$, then $F(\lambda)<0$ for any $\lambda_{-1}<\lambda<0$ and it is impossible to have negative roots of the equation $F(\lambda)=0$. The latter leads to the absence of unstable modes.
\item
Q-balls with $\omega=0$ are always classically unstable. The solution to equation \eqref{eqmatrixform}, describing the unstable mode, has the form
\begin{equation}
\Psi\sim\begin{pmatrix}\psi_{-1}(\vec x) \\ 0\end{pmatrix},\qquad \gamma=\sqrt{-\lambda_{-1}}.
\end{equation}
\item
It is clear that the proof presented above works for solutions of form \eqref{qballsolution} (and without nodes) in the space-time with compact spatial dimensions.
\item
The proof presented above {\em does not} work for the scalar condensate solutions of the form $f(\vec x)=f_{c}e^{i\omega t}$, where $f_{c}$ is a constant, even in the space-time with compact spatial dimensions (in this case the charge of the condensate is finite). Indeed, in our notations the classical stability condition \cite{Nugaev:2015rna} for such a condensate is
\begin{equation}\label{Sdefcondensate}
\frac{d^{2}V}{d(\phi^{*}\phi)^{2}}\biggl|_{\phi^{*}\phi=f_{c}^{2}}\ge 0,
\end{equation}
whereas
\begin{equation}
\omega^{2}=\frac{dV}{d(\phi^{*}\phi)}\biggl|_{\phi^{*}\phi=f_{c}^{2}}
\end{equation}
defines the value of $f_{c}$ \cite{Nugaev:2015rna}. With these relations, the operator $L_{+}$ has the form
\begin{equation}
L_{+}=-\Delta+2\frac{d^{2}V}{d(\phi^{*}\phi)^{2}}\biggl|_{\phi^{*}\phi=f_{c}^{2}}f_{c}^{2},
\end{equation}
which implies that, due to \eqref{Sdefcondensate}, $L_{+}$ has no negative eigenvalues. The latter makes the Vakhitov-Kolokolov approach inapplicable in this case. An extra indication of this fact is the absence of the corresponding translational mode ($\partial_{i}f_{c}\equiv 0$). Moreover, it is possible to show that for the stable scalar condensate satisfying \eqref{Sdefcondensate} the relation $\frac{dQ_{c}}{d\omega}>0$ fulfills \cite{Nugaev:2016wyt}.
\end{itemize}

\subsection{Two-field Q-balls}
Here we consider the action of form
\begin{equation}
S=\int dtd^{d}x\left(\partial_\mu\phi^*\partial^\mu\phi+\frac{1}{2}\partial_\mu\chi\partial^\mu\chi-V(\phi^*\phi,\chi)\right),
\end{equation}
where $\phi$ is a complex scalar field and $\chi$ is a real scalar field. For the two-field Q-ball, we consider the ansatz
\begin{eqnarray}
\phi(t,\vec x)=e^{i\omega t}f(r),\\
\chi(t,\vec x)=g(r),
\end{eqnarray}
where again $r=\sqrt{{\vec x}^{2}}$, $f(r)$ and $g(r)$ are real functions, $f(r)>0$ for any $r$, and
\begin{equation}
\partial_{r}f(r)|_{r=0}=0,\qquad \lim\limits_{r\to\infty}f(r)=0,\qquad \partial_{r}g(r)|_{r=0}=0,\qquad \lim\limits_{r\to\infty}g(r)=0.
\end{equation}
A well-known example of the soliton solution in such a model can be found in \cite{Friedberg:1976me}.

The functions $f(r)$ and $g(r)$ satisfy the equations of motion
\begin{eqnarray}\label{eqforf}
\omega^{2}f+\Delta f-\frac{\partial V}{\partial(\phi^{*}\phi)}\biggl|_{\substack{\phi^{*}\phi=f^{2}\\ \chi=g}}f=0,\\ \label{eqforg}
\Delta g-\frac{\partial V}{\partial\chi}\biggl|_{\substack{\phi^{*}\phi=f^{2}\\ \chi=g}}=0.
\end{eqnarray}
The charge of the Q-ball is defined by \eqref{qballcharge}.

We start from the ansatz
\begin{eqnarray}\label{pertdef2fields}
&&\phi(t,\vec x)=e^{i\omega t}f(r)+e^{i\omega t}e^{\gamma t}\left(u(\vec x)+iv(\vec x)\right),\\ \label{pertchidef2fields}
&&\chi(t,\vec x)=g(r)+e^{\gamma t}\varphi(\vec x),
\end{eqnarray}
where $\gamma$ is a real constant. Analogously to the case of the one-field Q-ball, the linearized equations of motion for the perturbations $u$, $v$, and $\varphi$ can be represented in the operator form
\begin{equation}
\begin{pmatrix}
L_{u} & Y & 0 \\
Y & L_{\varphi} & 0\\
0 & 0 & L_{-}
\end{pmatrix}\Psi=-\gamma^{2}\begin{pmatrix}u \\ \frac{1}{2}\varphi \\ 0\end{pmatrix}+2\omega\gamma\begin{pmatrix}v \\ 0 \\ -u\end{pmatrix},
\end{equation}
where
\begin{equation}
\Psi=\begin{pmatrix}u \\ \varphi \\ v\end{pmatrix},
\end{equation}
\begin{eqnarray}\label{Lu}
&&L_{u}=-\Delta+U(r)+2S(r)-\omega^{2},\\ \label{Lvarphi}
&&L_{\varphi}=-\frac{\Delta}{2}+W(r),\\
&&L_{-}=-\Delta+U(r)-\omega^{2}+\gamma^{2},
\end{eqnarray}
and
\begin{eqnarray}\label{U2field}
U(r)=\frac{\partial V}{\partial(\phi^{*}\phi)}\biggl|_{\substack{\phi^{*}\phi=f^{2}(r)\\ \chi=g(r)}},\qquad S(r)=\frac{\partial^{2}V}{\partial(\phi^{*}\phi)^{2}}\biggl|_{\substack{\phi^{*}\phi=f^{2}(r)\\ \chi=g(r)}}f^{2}(r),\\ \label{W2field}
W(r)=\frac{1}{2}\frac{\partial^{2}V}{\partial\chi^{2}}\biggl|_{\substack{\phi^{*}\phi=f^{2}(r)\\ \chi=g(r)}},\qquad
Y(r)=\frac{\partial^{2}V}{\partial(\phi^{*}\phi)\partial\chi}\biggl|_{\substack{\phi^{*}\phi=f^{2}(r)\\ \chi=g(r)}}f(r).
\end{eqnarray}
As in the one-field case, it is also possible to show that unstable modes can have only the form \eqref{pertdef2fields} and \eqref{pertchidef2fields}; see Appendix~B for details.

Let us define the matrix operators
\begin{equation}
\hat L=\begin{pmatrix}
L_{u} & Y & 0 \\
Y & L_{\varphi} & 0\\
0 & 0 & L_{-}
\end{pmatrix},\qquad L_{+}=
\begin{pmatrix}
L_{u} & Y \\
Y & L_{\varphi}
\end{pmatrix}.
\end{equation}

In full analogy with the case of one-field Q-balls, we can show that
\begin{enumerate}
\item
\begin{equation}
\expval{\hat L}{\Psi}=-\gamma^{2}\bra{u}\ket{u}-\frac{\gamma^{2}}{2}\bra{\varphi}\ket{\varphi}.
\end{equation}
\item
By multiplying the equation for the field $v$ by $f$, integrating the result over the spatial volume and using equation \eqref{eqforf}, we get
\begin{equation}
\bra{\Phi}\ket{\Psi}=0,\quad \textrm{where} \quad
\Phi=\begin{pmatrix}2\omega f \\ 0 \\ \gamma f\end{pmatrix}.
\end{equation}
Again, this relation is the consequence of the total charge conservation.
\item
By differentiating equations \eqref{eqforf} and \eqref{eqforg} with respect to $x^{i}$, we get
\begin{equation}
L_{+}\begin{pmatrix}\partial_{i}f \\ \partial_{i}g\end{pmatrix}=0,
\end{equation}
leading to
\begin{equation}
\hat L T_{i}=\hat L\begin{pmatrix}\partial_{i}f \\ \partial_{i}g \\ 0\end{pmatrix}=0.
\end{equation}
$T_{i}$ are the zero modes of the operator $\hat L$ such that $\bra{T_{i}}\ket{\Phi}=0$.
\item
By differentiating equations \eqref{eqforf} and \eqref{eqforg} with respect to $\omega$, we get
\begin{equation}
L_{+}\begin{pmatrix}\frac{df}{d\omega} \\ \frac{dg}{d\omega}\end{pmatrix}=\begin{pmatrix}2\omega f \\ 0\end{pmatrix},
\end{equation}
leading to
\begin{equation}\label{LPhieq2fields}
\hat L\begin{pmatrix}\frac{df}{d\omega} \\ \frac{dg}{d\omega} \\ \frac{f}{\gamma}\end{pmatrix}=\Phi.
\end{equation}
\end{enumerate}
Now we have all the ingredients which are necessary to obtain the classical stability criterion for two-field Q-balls. Again, since all the eigenstates of the operator $\hat L$ with the zero and negative eigenvalues are defined by the corresponding eigenstates of the operator $L_{+}$, we assume that there is only one negative eigenvalue of the operator $L_{+}$ (the same restriction on the number of negative eigenvalues of an analogous operator was obtained for the two-field model proposed and examined in \cite{Friedberg:1976me}). Analogously to the one-field case, it is not difficult to show that even if there exist other zero modes of the operator $L_{+}$, they do not contribute to the corresponding decomposition of $\Phi$. In this case we can perform all the necessary steps of the proof, in full analogy with the one-field case (i.e., considering the variational problem, expanding $\Psi$ and $\Phi$ in the eigenfunctions of the operator $\hat L$, constructing the function $F(\lambda)$, etc.). Finally, we get
\begin{equation}\label{F0deftwofield}
F(0)=\expval{{\hat L}^{-1}}{\Phi}=\int \begin{pmatrix}2\omega f & 0 & \gamma f\end{pmatrix}\begin{pmatrix}\frac{df}{d\omega} \\ \frac{dg}{d\omega} \\ \frac{f}{\gamma}\end{pmatrix}d^{d}x=
\int \left(2\omega f\frac{df}{d\omega}+f^{2}\right)d^{d}x=\frac{1}{2}\frac{dQ}{d\omega}.
\end{equation}
The latter means that there are no unstable modes with $\gamma\neq 0$ for $\frac{dQ}{d\omega}<0$.

It is possible that the positive part of the spectrum of the operator $L_{+}$ is purely continuous and starts from $\lambda_{c}=0$. In this case we expect $d(\tilde\lambda)\to 0$ for $\tilde\lambda\to 0$ because of \eqref{LPhieq2fields}, which makes $\frac{dQ}{d\omega}$ finite. Such a situation may appear, for example, in the two-field models discussed in \cite{Levin:2010gp,Nugaev:2016uqd}. Again, in full analogy with the one-field case (see \eqref{onlycontspectrum}), it is easy to show that in this case there are no unstable modes for $\frac{dQ}{d\omega}<0$ (of course, if $L_{+}$ has only one negative eigenvalue).

We have shown that the Vakhitov-Kolokolov method \cite{VK,Kolokolov} for obtaining the classical stability criterion for the systems described by the nonlinear Schr\"{o}dinger equation can be generalized not only to the relativistic case of one-field Q-balls, but also to the case of, say, two-field Q-balls. The only substantial difference of the two-field case with the one-field case is the form of the operator $L_{+}$: in the one-field case it has the standard ``quantum mechanical'' form, whereas in the two-field case it has the matrix form with diagonal entries having the form of ``quantum mechanical'' operators and with the nonzero off-diagonal elements (like the one in \cite{Friedberg:1976me}). In fact, the eigenvalue problem for the operator $L_{+}$ is reduced to the system of coupled second-order differential equations of motion. However, due to the fact that the diagonal entries of this matrix are the operators which are bounded from below, even though there exist nonzero off-diagonal elements, one can hope that at least in the most cases the whole operator $L_{+}$ is bounded from below and has only one negative eigenvalue.

Now we are ready to pass to the case of $U(1)$ gauged Q-balls.

\section{Classical (in)stability of U(1) gauged Q-balls}

It is reasonable to suppose that if the same classical stability criterion ($\frac{dQ}{d\omega}<0$) is valid for $U(1)$ gauged Q-balls, then it can be obtained in the way fully analogous to the one used for ordinary Q-balls. As will be shown below, all the technical steps, which are necessary for obtaining the proof for $U(1)$ gauged Q-balls, can be performed in the same way as for ordinary Q-balls. However, in the gauged case there exist some obstacles, which invalidate the classical stability criterion $\frac{dQ}{d\omega}<0$. This conclusion will be supported by the numerical simulations in several explicit cases.

Let us consider the action of the form
\begin{equation}\label{actiongauged}
S=\int
dtd^3x\left((\partial^{\mu}\phi^{*}-ieA^{\mu}\phi^{*})(\partial_{\mu}\phi+ieA_{\mu}\phi)-V(\phi^{*}\phi)-\frac{1}{4}F_{\mu\nu}F^{\mu\nu}\right),
\end{equation}
where $e$ is the coupling constant. We will focus only on the $(3+1)$-dimensional case, for which the $A_{0}$ component of the gauge field of the Q-ball behaves as $A_{0}(r)\sim\frac{1}{r}$ at large $r$. We take the standard spherically symmetric ansatz for the scalar and gauge fields \cite{Rosengauged,Lee:1988ag}:
\begin{eqnarray}\label{ans1}
\phi(t,\vec x)&=&\textrm{e}^{i\omega t}f(r),\qquad
f(r)|_{r\to\infty}\to 0, \qquad \partial_{r}f(r)|_{r=0}=0,\\
\label{ans2}A_{0}(t,\vec x)&=&A_{0}(r),\qquad\,\,
A_{0}(r)|_{r\to\infty}\to 0,\qquad
\partial_{r}A_{0}(r)|_{r=0}=0,\\ \label{ans3} A_{i}(t,\vec
x)&\equiv &0,
\end{eqnarray}
where again $r=\sqrt{\vec x^{2}}$, $f(r)>0$, and $A_{0}(r)$ are real
functions. The equations of motion for $f(r)$ and $A_{0}(r)$, following from action
\eqref{actiongauged}, take the form
\begin{eqnarray}\label{eqg1}
\Delta A_{0}-2e(\omega+eA_{0})f^2=0,\\
\label{eqg2} (\omega+eA_{0})^2f+\Delta f-\frac{dV}{d(\phi^{*}\phi)}\biggl|_{\phi^{*}\phi=f^{2}}f=0.
\end{eqnarray}
The charge of the Q-ball is defined by
\begin{equation}\label{qballchargegauged}
Q=2\int (\omega+eA_{0})f^{2}d^{3}x.
\end{equation}

\subsection{Analytical considerations}
For perturbations above the Q-ball solution, from the very beginning we consider the following ansatz:
\begin{eqnarray}\label{instgauge1}
&&\phi(t,\vec x)=e^{i\omega t}f(r)+e^{i\omega t}e^{\gamma t}\left(u(\vec x)+iv(\vec x)\right),\\
&&A_{0}(t,\vec x)=A_{0}(r)+e^{\gamma t}a_{0}(\vec x),\\ \label{instgauge3}
&&A_{i}(t,\vec x)=e^{\gamma t}a_{i}(\vec x)
\end{eqnarray}
with $\gamma\neq 0$. The corresponding linearized equations of motion for the fields $u$, $v$, $a_{0}$, and $a_{i}$ take the form (we do not present the derivation here --- it is straightforward, although rather bulky)
\begin{eqnarray}\label{eqgauged1}
\Delta u-Uu-2Su+(\omega+eA_{0})^{2}u-\gamma^{2}u+2(\omega+eA_{0})\gamma v+2e(\omega+eA_{0})fa_{0}=0,\\
\Delta v-Uv+(\omega+eA_{0})^{2}v-\gamma^{2}v-2(\omega+eA_{0})\gamma u-e\gamma f a_{0}+ef\partial_{i}a_{i}+2e\partial_{i}f a_{i}=0,\\
\Delta a_{0}-2e^{2}f^{2}a_{0}-\gamma\partial_{i}a_{i}-4e(\omega+eA_{0})fu-2e\gamma fv=0,\\ \label{eqgauged4}
\Delta a_{i}-\partial_{i}\partial_{j}a_{j}-2e^{2}f^{2}a_{i}+\gamma\partial_{i}a_{0}-\gamma^{2}a_{i}-2e(f\partial_{i}v-v\partial_{i}f)=0,
\end{eqnarray}
where $U(r)$ and $S(r)$ are defined by \eqref{UGdef}. From here and below the terms like $\partial_{i}a_{i}$ or $\partial_{i}f a_{i}$ should be considered as $\partial_{i}a_{i}=\sum\limits_{k=1}^{3}\partial_{k}a_{k}$ and $\partial_{i}f a_{i}=\sum\limits_{k=1}^{3}a_{k}\partial_{k}f$, respectively. One can see that the resulting system of linearized equations of motion \eqref{eqgauged1}--\eqref{eqgauged4} for $u$, $v$, $a_{0}$, and $a_{i}$ is time independent, which is expected for the method of separation of variables. The parameter $\gamma$ in \eqref{instgauge1}--\eqref{instgauge3} is the same for the scalar and gauge fields, otherwise we would get extra time-dependent terms in the linearized equations for $u$, $v$, $a_{0}$, and $a_{i}$, which would make the Vakhitov-Kolokolov approach unusable.

Equations \eqref{eqgauged1}--\eqref{eqgauged4} are invariant under the transformations
\begin{equation}
v\to v+ef\beta(\vec x),\qquad a_{0}\to a_{0}-\gamma\beta(\vec x),\qquad a_{i}\to a_{i}-\partial_{i}\beta(\vec x),
\end{equation}
which are the consequence of the gauge transformations $\phi\to\phi e^{ie\alpha}$, $A_{\mu}\to A_{\mu}-\partial_{\mu}\alpha$ with $\alpha(t,\vec x)=e^{\gamma t}\beta(\vec x)$. It is convenient to work with gauge invariant variables, so we introduce the new gauge invariant fields
\begin{equation}\label{gaugeinvfields}
\xi=v+ef\frac{\partial_{i}a_{i}}{\Delta},\qquad q_{0}=a_{0}-\gamma\frac{\partial_{i}a_{i}}{\Delta},\qquad q_{i}=a_{i}-\partial_{i}\left(\frac{\partial_{j}a_{j}}{\Delta}\right),
\end{equation}
where $\frac{(\dots)}{\Delta}=\Delta^{-1}(\dots)$ with $\Delta^{-1}$ being the inverse Laplace operator. In these notations, equations of motion \eqref{eqgauged1}--\eqref{eqgauged4} take the form
\begin{eqnarray}\label{eqgauged1a}
L_{u}u-2e(\omega+eA_{0})fq_{0}&=&-\gamma^{2}u+2(\omega+eA_{0})\gamma\xi,\\ \label{eqgauged2a}
L_{q}q_{0}-2e(\omega+eA_{0})fu&=&e\gamma f\xi,\\ \label{eqgauged3a}
L_{-}\xi&=&-2(\omega+eA_{0})\gamma u-e\gamma f q_{0}+2e\partial_{i}f q_{i},\\ \label{eqgauged4a}
L_{q}q_{i}-\frac{\gamma^{2}}{2}q_{i}&=&-\frac{\gamma}{2}\partial_{i}q_{0}+e(f\partial_{i}\xi-\xi\partial_{i}f),
\end{eqnarray}
where
\begin{eqnarray}
&&L_{u}=-\Delta+U(r)+2S(r)-(\omega+eA_{0})^{2},\\ \label{Lqdef}
&&L_{q}=\frac{\Delta}{2}-e^{2}f^{2},\\
&&L_{-}=-\Delta+U(r)-(\omega+eA_{0})^{2}+\gamma^{2}.
\end{eqnarray}
Equations \eqref{eqgauged1a}--\eqref{eqgauged4a} suggest the following form of the operators $\hat L$ and $L_{+}$:
\begin{equation}\label{hatLgaugeddef}
\hat L=\begin{pmatrix}
L_{u} & -2e(\omega+eA_{0})f & 0 & 0 \\
-2e(\omega+eA_{0})f & L_{q} & 0 & 0 \\
0 & 0 & L_{-} & 0 \\
0 & 0 & 0 & \left(L_{q}-\frac{\gamma^{2}}{2}\right)I_{3\times 3}
\end{pmatrix}
\end{equation}
and
\begin{equation}\label{Lplusgaugeddef}
L_{+}=
\begin{pmatrix}
L_{u} & -2e(\omega+eA_{0})f \\
-2e(\omega+eA_{0})f & L_{q}
\end{pmatrix},
\end{equation}
where $I_{3\times 3}$ is the $3\times 3$ unit matrix. In these notations, equations \eqref{eqgauged1a}--\eqref{eqgauged4a} can be rewritten as
\begin{equation}
\hat L\Psi=\begin{pmatrix}
-\gamma^{2}u+2(\omega+eA_{0})\gamma\xi \\ e\gamma f\xi \\ -2(\omega+eA_{0})\gamma u-e\gamma f q_{0}+2e\partial_{i}f q_{i} \\
-\frac{\gamma}{2}\partial_{1}q_{0}+e(f\partial_{1}\xi-\xi\partial_{1}f) \\ -\frac{\gamma}{2}\partial_{2}q_{0}+e(f\partial_{2}\xi-\xi\partial_{2}f) \\
-\frac{\gamma}{2}\partial_{3}q_{0}+e(f\partial_{3}\xi-\xi\partial_{3}f)
\end{pmatrix},\qquad \Psi=\begin{pmatrix}
u \\ q_{0} \\ \xi \\
q_{1} \\ q_{2} \\ q_{3}
\end{pmatrix}.
\end{equation}
As in the case of ordinary (nongauged) Q-balls, we can show that
\begin{enumerate}
\item
\begin{equation}\label{expvalhatL}
\expval{\hat L}{\Psi}=-\gamma^{2}\bra{u}\ket{u}.
\end{equation}
In derivation of this relation we have used the integration by parts in terms like $\bra{q_{i}}{\partial_{i}}\ket{q_{0}}$ and the fact that $\partial_{i}q_{i}=0$; see the definition of $q_{i}$ in \eqref{gaugeinvfields}.
\item
By multiplying equation \eqref{eqgauged3a} by $f$, integrating the result over the spatial volume and using equation \eqref{eqg2} and $\partial_{i}q_{i}=0$, we get
\begin{equation}
\bra{\Phi}\ket{\Psi}=0,\quad \textrm{where} \quad
\Phi=\begin{pmatrix}2(\omega+eA_{0})f \\ ef^{2} \\ \gamma f \\ 0 \\ 0 \\ 0\end{pmatrix}.
\end{equation}
Again, this relation is also the consequence of the total charge conservation.
\item
By differentiating equations \eqref{eqg1} and \eqref{eqg2} with respect to $x^{i}$, we get
\begin{equation}
L_{+}\begin{pmatrix}\partial_{i}f \\ \partial_{i}A_{0}\end{pmatrix}=0,
\end{equation}
leading to
\begin{equation}\label{translmodesgauged}
\hat L T_{i}=\hat L\begin{pmatrix}\partial_{i}f \\ \partial_{i}A_{0} \\ 0 \\ 0 \\ 0 \\ 0\end{pmatrix}=0.
\end{equation}
Here $T_{i}$ are the zero modes of the operator $\hat L$ such that $\bra{T_{i}}\ket{\Phi}=0$.
\item
By differentiating equations \eqref{eqg1} and \eqref{eqg2} with respect to $\omega$, we get
\begin{equation}\label{omegamodesgauged}
L_{+}\begin{pmatrix}\frac{df}{d\omega} \\ \frac{dA_{0}}{d\omega}\end{pmatrix}=\begin{pmatrix}2(\omega+eA_{0})f \\ ef^{2}\end{pmatrix},
\end{equation}
leading together with $L_{-}f=\gamma^{2} f$ to
\begin{equation}
\hat L\begin{pmatrix}\frac{df}{d\omega} \\ \frac{dA_{0}}{d\omega} \\ \frac{f}{\gamma} \\ 0 \\ 0 \\ 0\end{pmatrix}=\Phi.
\end{equation}
\end{enumerate}
Thus, we have all the key ingredients which are necessary for performing the proof along the lines of the proof for nongauged Q-balls.

A few important remarks are in order here.
\begin{itemize}
\item
For the spherically symmetric perturbations $q_{i}\equiv 0$. Indeed, in such a case $a_{i}(\vec x)=x_{i}a(r)$, which leads to $q_{i}\equiv 0$ for \eqref{gaugeinvfields}.
\item
It should be noted that equations \eqref{eqgauged2a}--\eqref{eqgauged4a} are not independent. Indeed, by acting $\partial_{i}$ on equation \eqref{eqgauged4a} and using \eqref{eqgauged3a} we can get equation \eqref{eqgauged2a} and, alternatively, by acting $\partial_{i}$ on equation \eqref{eqgauged4a} and using \eqref{eqgauged2a} we can get equation \eqref{eqgauged3a}. This suggests that one can simply remove equation \eqref{eqgauged2a} or equation \eqref{eqgauged3a} from the whole system of equations. However, at least for performing the derivation of the stability criterion along the lines of the Vakhitov-Kolokolov approach, all four equations of motion are necessary. Indeed, without equation \eqref{eqgauged2a} it is impossible to get the relations \eqref{translmodesgauged} and \eqref{omegamodesgauged}; whereas without equation \eqref{eqgauged2a} or equation \eqref{eqgauged3a} we cannot get the relation
$$
\expval{\hat L}{\Psi}=-(\textrm{some nonnegative value}),
$$
for the matrix operator $\hat L$, which is crucial for the proof. For the nonspherically symmetric perturbations the only obvious exception is $e=0$. In this case, equations for $q_{0}$, $q_{i}$ completely decouple ($A_{0}(r)\equiv 0$ in this case) and we can use the equations for the fields $u$ and $\xi$ only --- effectively the system reduces to the one-field case discussed in Subsection~2.1.
\end{itemize}

However, the operator $\hat L$ defined by \eqref{hatLgaugeddef} is unbounded from below. This follows from the fact that the operator $L_{q}-\frac{\gamma^{2}}{2}$ is unbounded from below (look at the sign in front of $\Delta$ in \eqref{Lqdef}), whereas the corresponding off-diagonal elements are equal to zero. Thus, the corresponding negative part of the spectrum is defined only by the spectrum of the operator $L_{q}-\frac{\gamma^{2}}{2}$. Moreover, due to the fact that $f(r)$ rapidly tends to zero at $r\to\infty$, one expects that the operator $L_{q}-\frac{\gamma^{2}}{2}$ has a negative continuous spectrum starting at $-\gamma^{2}$. In this situation the sign of the corresponding function $F(0)$ does not provide any information about possible classical stability regions of the Q-ball, at least within the framework of the generalized Vakhitov-Kolokolov approach.

At the moment it is not clear whether or not the existence of the negative spectrum is just a technical artifact of the approach used here; we can only guess. Maybe, due to the repulsive nature of the electromagnetic field, this means that nonspherically symmetric perturbations (i.e., such that $q_{i}\not\equiv 0$) destroy any $U(1)$ gauged Q-ball.

Just for completeness of the study, we can proceed with considering only the spherically symmetric perturbations above the gauged Q-ball. In this case $q_{i}\equiv 0$, so we can use the set of reduced operators and fields to examine this problem. Namely, we take
\begin{equation}\label{hatLgaugeddefred}
\hat L_{3}=\begin{pmatrix}
L_{+} & 0\\
0 & L_{-}
\end{pmatrix},\qquad \Psi_{3}=\begin{pmatrix}
u \\ q_{0} \\ \xi
\end{pmatrix},\qquad \Phi_{3}=\begin{pmatrix}2(\omega+eA_{0})f \\ ef^{2} \\ \gamma f\end{pmatrix}.
\end{equation}
In this case
\begin{equation}
\hat L_{3}\begin{pmatrix}\partial_{i}f \\ \partial_{i}A_{0} \\ 0\end{pmatrix}=0,\qquad \hat L_{3}\begin{pmatrix}\frac{df}{d\omega} \\ \frac{dA_{0}}{d\omega} \\ \frac{f}{\gamma} \end{pmatrix}=\Phi_{3},\qquad \expval{\hat L_{3}}{\Psi_{3}}=-\gamma^{2}\bra{u}\ket{u}.
\end{equation}
We see that this setup is almost the same as the one in the two-field case discussed in Subsection~2.2. So, performing the same steps as those in Section~2, we arrive at
\begin{equation}
F(0)=\expval{{\hat L_{3}}^{-1}}{\Phi_{3}}=\int \left(2(\omega+eA_{0})f\frac{df}{d\omega}+ef^{2}\frac{dA_{0}}{d\omega}+f^{2}\right)d^{3}x=
\frac{1}{2}\frac{dQ}{d\omega},
\end{equation}
where the gauged Q-ball charge is defined by \eqref{qballchargegauged}. Thus, one may think that if $\frac{dQ}{d\omega}<0$, then equation \eqref{expvalhatL} cannot be fulfilled for a nonzero $\Psi$, leading to the absence of classical instabilities.

Formally, we have obtained the classical stability criterion for the spherically symmetric perturbations above the gauged Q-ball solution. However, the question about the number of negative eigenvalues of the operator $L_{+}$, defined by \eqref{Lplusgaugeddef}, appears to be rather complicated, even though for the spherically symmetric perturbations it effectively reduces to the one-dimensional problem. Indeed, the diagonal entry $L_{u}$ of the operator $L_{+}$ is bounded from below, whereas the diagonal entry $L_{q}$ is {\em unbounded} from below. This differs considerably from the case of the ordinary two-field Q-ball, in which all the diagonal entries are bounded from below.

Let us look at the eigenvalue problem for the operator $L_{+}$ more precisely. For $r\to\infty$, using the fact that $f(r)$ very rapidly tends to zero with $r$, we get
\begin{equation}
\frac{\Delta}{2}q_{0}\approx\lambda q_{0}.
\end{equation}
It means that for any $\lambda<0$ and for $r\to\infty$
\begin{equation}
q_{0}\sim\frac{\sin(\sqrt{-2\lambda}\,r)}{r},\qquad q_{0}\sim\frac{\cos(\sqrt{-2\lambda}\,r)}{r}.
\end{equation}
Such a behavior of the solutions corresponds to the modes from continuous spectrum. Thus, even if there exists a solution to the eigenvalue problem of the operator $L_{+}$ with the negative eigenvalue $\lambda$, it belongs to the continuous spectrum, which invalidates the use of the Vakhitov-Kolokolov approach in this case. Of course, there are nonzero off-diagonal elements of the operator $L_{+}$, so one can assume that at least in some very special cases the spectrum of the operator $L_{+}$ does not contain negative eigenvalues at all, demonstrating the classical stability of the corresponding gauged Q-ball with respect to spherically symmetric perturbations. But so far we have no definite answer to the question of whether or not such a possibility can be realized.

Note that, contrary to the cases discussed in Subsections~2.1 and 2.2, here we do not have a rigorous mathematical proof that the ansatz \eqref{instgauge1}--\eqref{instgauge3} is the only possible for unstable modes. However, it is the simplest form of perturbations for unstable modes that passes through the linearized equations of motion; so even if there exist other types of unstable modes, the use of the ansatz \eqref{instgauge1}--\eqref{instgauge3} is sufficient for demonstration of the inapplicability of the Vakhitov-Kolokolov approach for gauged Q-balls.

In principle, the fact that the Vakhitov-Kolokolov approach does not work for gauged Q-balls does not mean that the case in which gauged Q-balls with $\frac{dQ}{d\omega}<0$ are classically stable and gauged Q-balls with $\frac{dQ}{d\omega}>0$ are classically unstable cannot be realized, such a conclusion does not follow from the results presented above. However, these results imply that, contrary to the case of nongauged Q-balls, one cannot suppose that in the most cases there is a direct connection between the sign of $\frac{dQ}{d\omega}$ and the classical stability of gauged Q-balls. In the next subsection we will demonstrate that it is indeed so.

\subsection{Numerical simulations}
In the present section we will examine the classical stability or
instability of gauged Q-balls by performing numerical simulations
in the spherically symmetric case. We will consider the models
with the following scalar field potentials:
\begin{eqnarray}
\label{potential1}
V(\phi^{*}\phi)&=&M^{2}\phi^{*}\phi-\lambda_{\phi}(\phi^{*}\phi)^{2},\\
\label{potential2}
V(\phi^{*}\phi)&=&M^2\phi^{*}\phi\,\theta\left(1-\frac{\phi^{*}\phi}
{v^2}\right)+M^2v^2\theta\left(\frac{\phi^{*}\phi}{v^2}-1\right),\\
\label{potential3}
V(\phi^{*}\phi)&=&-\mu^2\phi^*\phi\ln(\beta^2\phi^*\phi),
\end{eqnarray}
where $\lambda_{\phi}>0$, $\theta$ is the Heaviside step function
with the convention $\theta(0)=\frac{1}{2}$. $M$, $v$, $\mu$ and
$\beta$ are the parameters of the scalar field potentials.

For the case of nongauged Q-balls, potential \eqref{potential1}
was introduced in \cite{Anderson:1970et}. The piecewise potential
\eqref{potential2} was introduced in a more general form in
\cite{Rosen0}, the corresponding nongauged Q-balls in such a
model were thoroughly examined in \cite{Gulamov:2013ema}
(see also \cite{Theodorakis:2000bz} for a model with similar
scalar field potential). Potential \eqref{potential3} was introduced in
\cite{Rosen1}; the corresponding nongauged Q-balls in this model
were thoroughly examined in \cite{MarcVent}. An interesting feature
of the nongauged theory with potential \eqref{potential3} is that
not only can the Q-ball solutions be obtained analytically in this
model, but also the analysis of perturbations above the Q-ball can
be made fully analytically \cite{MarcVent}, providing an explicit
demonstration of the validity of the classical stability criterion
$\frac{dQ}{d\omega}<0$ in this model.

For the $U(1)$ gauged Q-balls, potential \eqref{potential1} was
considered in \cite{Gulamov:2015fya}, potential \eqref{potential2}
--- in \cite{Gulamov:2015fya} and (in a more general form) in
\cite{Gulamov:2013cra}, and potential \eqref{potential3} --- in
\cite{Dzhunushaliev:2012zb,Tamaki:2014oha,Gulamov:2013cra}

Now we turn to a description of our method for examining classical (in)stability
of gauged Q-balls. Since we will consider only the spherically symmetric evolution with $A_{\varphi}\equiv 0$, $A_{\theta}\equiv 0$, from the very beginning we impose the gauge
\begin{equation}
A_{r}(t,r)\equiv 0.
\end{equation}
In this case the equations of motion take the form
\begin{eqnarray}
\label{numeq1}
\partial_{t}^{2}\phi+2ieA_{0}\partial_{t}\phi+ie\partial_{t}A_{0}\phi-
e^{2}A_{0}^{2}\phi-\frac{1}{r}\partial_{r}^{2}(r\phi)+
\frac{dV}{d(\phi^{*}\phi)}\phi=0,\\
\label{numeq2}
\frac{1}{r}\partial_{r}^{2}(rA_{0})+ie\left(\phi^{*}\partial_{t}\phi-\phi\partial_{t}\phi^{*}+
2ieA_{0}\phi^{*}\phi\right)=0.
\end{eqnarray}
There is also the equation of motion for the field $A_{r}$, which
plays the role of a constraint in the gauge $A_{r}\equiv 0$. However, this equation can be deduced from equations \eqref{numeq1} and
\eqref{numeq2}, so we keep only equations \eqref{numeq1} and \eqref{numeq2} for the numerical simulations. We will be interested in spatially localized and regular configurations satisfying the boundary conditions
\begin{eqnarray}
\label{bc1}
\partial_{r}\phi(t,r)|_{r=0}=0,\qquad \phi(t,r)|_{r\to\infty}=0,&&\\
\label{bc2}
\partial_{r}A_{0}(t,r)|_{r=0}=0,\qquad  A_{0}(t,r)|_{r\to\infty}=0.&&
\end{eqnarray}

We begin with finding numerically gauged Q-ball solutions
$f(r)$ and $A_0(r)$, satisfying equations~\eqref{eqg1} and \eqref{eqg2}
with the boundary conditions~\eqref{ans1} and \eqref{ans2}, using the shooting method.
Then we slightly perturb the obtained solutions for the scalar field as
\begin{eqnarray}
\label{in1}
\phi(t,r)|_{t=0}&=&f(r)+\delta \phi(r),\\
\label{in2}
{\partial_{t}\phi}(t,r)|_{t=0}&=&i \omega f(r) +\delta {\dot \phi}(r),
\end{eqnarray}
where $\delta \phi(r)$, $\delta {\dot \phi}(r)$ are small
perturbations. These perturbations are generated randomly in the Fourier
space with the white noise spectrum; their average amplitude is of the order
of $10^{-2}f_{max}$ in the region of the gauged Q-ball core, where $f_{max}$ is the maximum value of the scalar field profile of a Q-ball.
Then we numerically evolve these perturbed configurations forward in time. For a gauged
Q-ball solution, we make $10$ different attempts to randomly perturb it and evolve it forward in time.
In our simulations, we use the stable second-order iterated Crank-Nicolson scheme \cite{CrankNicolson} (see Appendix~C for details).
Such perturbations eventually destroy classically unstable gauged Q-balls, while the
classically stable gauged Q-balls survive and drop some charge outside its core by means of spherical
waves.

We start with gauged Q-balls in a theory with potential \eqref{potential1}. Our simulations
show that there are no classically stable gauged Q-ball solutions in this theory. An analogous result was obtained
for the corresponding nongauged Q-balls in \cite{Anderson:1970et}. Formally, this fact is in agreement with the classical stability criterion $\frac{dQ}{d\omega}<0$, because all Q-balls in this model are such that $\frac{dQ}{d\omega}>0$ in the nongauged \cite{Anderson:1970et} and in the gauged \cite{Gulamov:2015fya} cases.

\begin{figure}[!ht]
\centering
\includegraphics[width=0.8\textwidth]{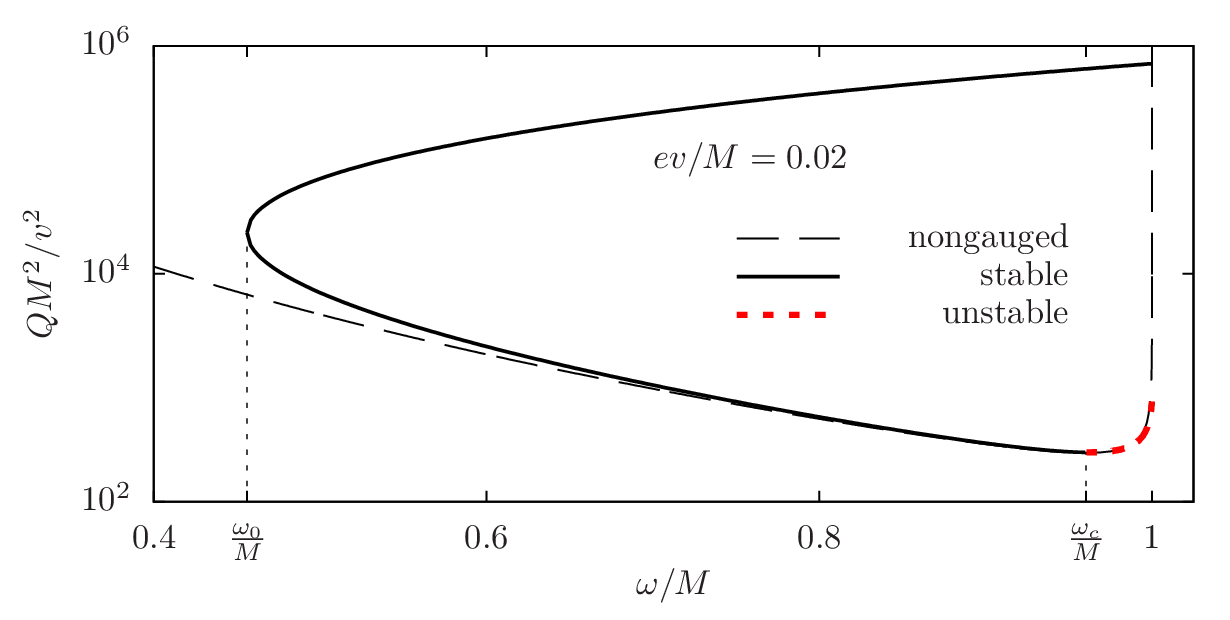}
\caption{$Q(\omega)$ dependence of gauged Q-balls in the model with potential \eqref{potential2} for $\frac{ev}{M}=0.02$. The long-dashed line stands for the nongauged Q-balls. The solid line stands for the classically stable gauged Q-balls. The short-dashed line stands for the classically unstable gauged Q-balls.
$\omega_0$ and $\omega_c$ correspond to the points where $\frac{dQ}{d\omega}=\infty$ and $\frac{dQ}{d\omega}=0$, respectively. From here and below, we present only the case $\omega\ge 0$, the case $\omega<0$ is fully symmetric.}
\label{figure3}
\end{figure}

Now we turn to the model with potential \eqref{potential2}. According to our simulations with different values of the effective coupling constant $\frac{ev}{M}$ (for the details of the model, as well as for the $Q(\omega)$ dependencies for different values of $\frac{ev}{M}$, see \cite{Gulamov:2015fya}), there exist both classically stable and unstable gauged Q-balls. A typical example is presented in Fig.~\ref{figure3}.\footnote{Note that, contrary to the case of ordinary Q-balls, gauged Q-balls with finite charge and energy exist even for $\omega=M$ in the model with potential \eqref{potential2}; see \cite{Gulamov:2015fya}. It is interesting to note that it is the parameter $\omega+eA_{0}(0)$ that uniquely characterizes a gauged Q-ball in this model \cite{Gulamov:2015fya}, not $\omega$ like in the nongauged case.} One can see that the derivative $\frac{dQ}{d\omega}$ can be positive and negative; it changes its sign in two points $\omega_0$, $\omega_c$ which correspond to $\frac{dQ}{d\omega}=\infty$ (for details, see \cite{Gulamov:2015fya}) and $\frac{dQ}{d\omega}=0$, respectively. Classically unstable gauged Q-balls are marked by the short-dashed line in Fig.~\ref{figure3}. For these solutions the inequality $\frac{dQ}{d\omega}>0$ fulfills, which formally is in agreement with the classical stability criterion. Meanwhile, the inequality $\frac{dQ}{d\omega}>0$ also fulfills for gauged Q-balls on the upper branch of the $Q(\omega)$ dependence. We have failed to find any instabilities of such gauged Q-balls during our numerical simulations with different values of $\frac{ev}{M}$. Note that this fact does not contradict the classical stability criterion: first, we have examined only the case of spherically symmetric perturbations, whereas an unstable mode (if exists) for these solutions can be nonspherically symmetric. Second, as it was mentioned above, even in the case of ordinary (nongauged) Q-balls, the classical stability criterion does not guarantee that there exist unstable modes in the case $\frac{dQ}{d\omega}>0$, it just states that there are no classical instabilities if $\frac{dQ}{d\omega}<0$. On the other hand, this situation differs from the case of nongauged Q-balls, for which usually there exist unstable modes for Q-balls with $\frac{dQ}{d\omega}>0$.

Now let us turn to the most complicated case of gauged Q-balls in the model with potential \eqref{potential3}. This model was discussed in detail in \cite{Tamaki:2014oha}, demonstrating rather complicated structure of the gauged Q-ball solutions. Contrary to the cases of potentials \eqref{potential1} and \eqref{potential2}, here we will not discuss all types of the gauged Q-ball solutions which exist in this model. We will restrict ourselves only to solutions which will be useful for examining classical stability.

We start with the values of the gauge coupling constant such that $e^{2}\ll\beta^{2}\mu^{2}$ holds, corresponding to the case in which the gauge interaction is much weaker than the scalar field self-interaction. In such a case the characteristics of gauged Q-balls are very close to those of the corresponding nongauged Q-balls \cite{Gulamov:2013cra}. Numerical simulations with $\frac{e}{\beta\mu}=0.1$ show that gauged Q-balls with $\frac{dQ}{d\omega}<0$ are classically stable with respect to spherically symmetric perturbations, whereas gauged Q-balls with $\frac{dQ}{d\omega}>0$ are classically unstable with respect to spherically symmetric perturbations. This result is in agreement with the classical stability criterion for nongauged Q-balls. One may assume that it happens because the backreaction of the gauge field is very small and classical stability or instability is governed mainly by the scalar field.

Now we turn to the values of the gauge coupling constant such that $e^{2}\gtrsim\beta^{2}\mu^{2}$ holds, corresponding to the case in which
the gauge interaction becomes comparable with the scalar field self-interaction. For the numerical analysis, we take $\frac{e}{\beta\mu}=1.1$. In this case, there exist many families of the gauged Q-ball solutions, satisfying equations \eqref{eqg1} and \eqref{eqg2} with the boundary conditions \eqref{ans1} and \eqref{ans2}. These solutions describe gauged Q-balls with a different and rather nontrivial form.
\begin{figure}[!ht]
\begin{tabular}{cc}
\includegraphics[width=0.47\linewidth]{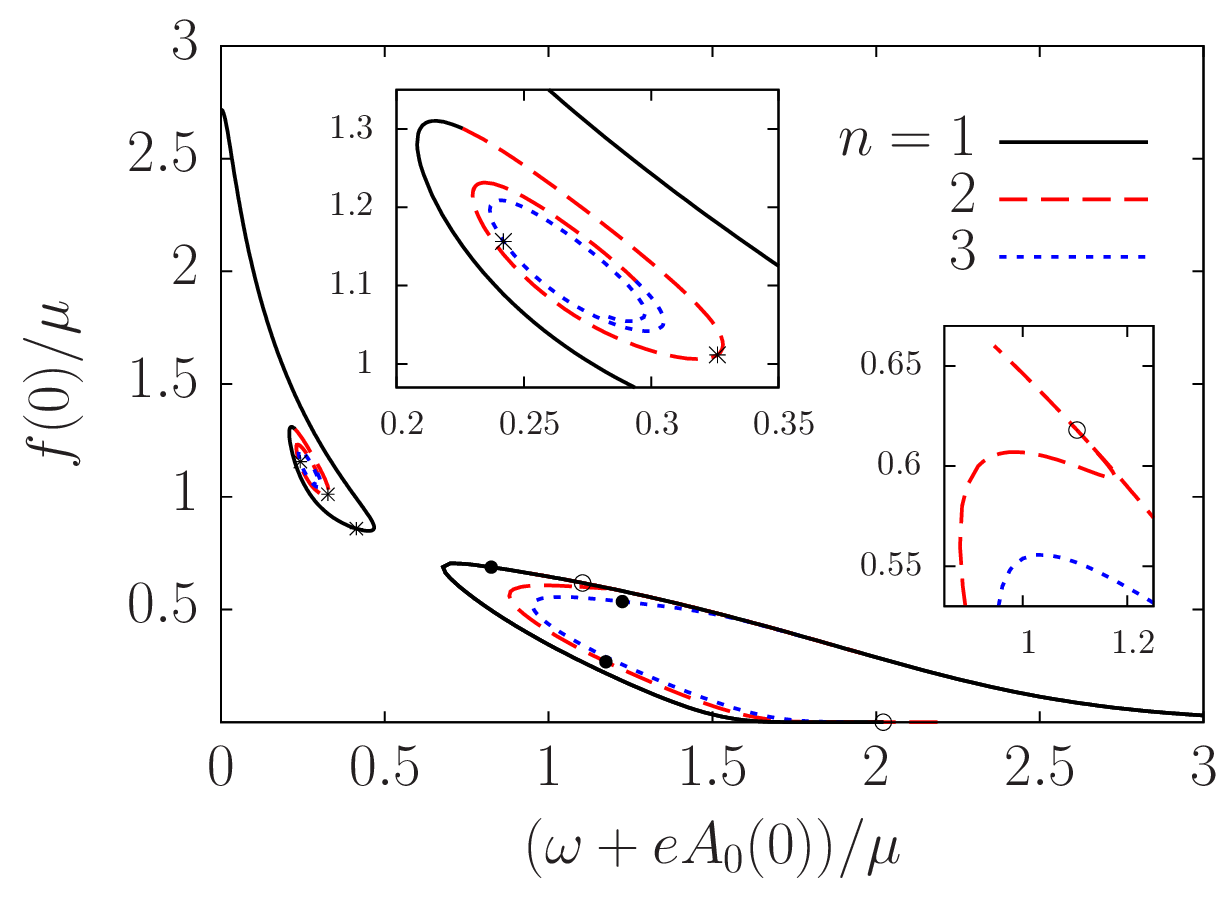}&
\includegraphics[width=0.47\linewidth]{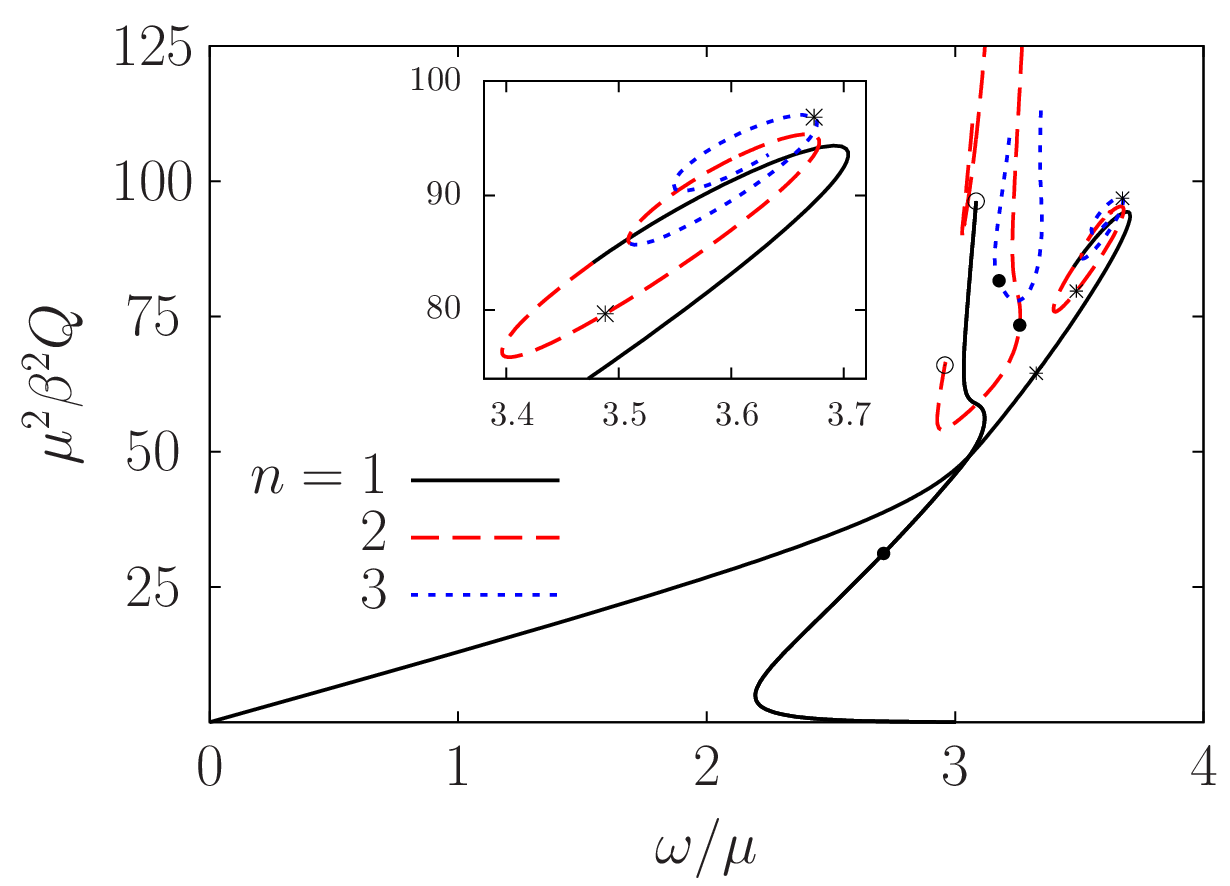}\\
$(a)$&$(b)$\\
\end{tabular}
\caption{Gauged Q-balls in the model with potential \eqref{potential3} for ${e}/{\beta\mu}=1.1$ --- the plane of initial data $(f(0),\,\omega+eA_0(0))$ $(a)$ and $Q(\omega)$ dependence $(b)$. Different families of solutions are represented by different lines. On the main part of plot $(a)$, some parts of the curves, corresponding to the lines which are visible on plot $(b)$, appear to be hidden by the solid line on plot $(a)$ (this happens because some solutions of different types have the values of initial data, which are much closer to each other than the effective resolution of plot $(a)$). In particular, the upper circle on plot $(a)$ lies on the long-dashed line, which is hidden by the solid curve. Some of such hidden lines are shown in the right inset of plot $(a)$, where only the lines for solutions with $n=2,3$ are presented. The scalar field profiles of the solutions marked by asterisks, dots and circles are presented in Fig.~\ref{figure5}.}
\label{figure4}
\end{figure}
\begin{figure}[!ht]
\begin{tabular}{ccc}
\includegraphics[width=0.30\linewidth]{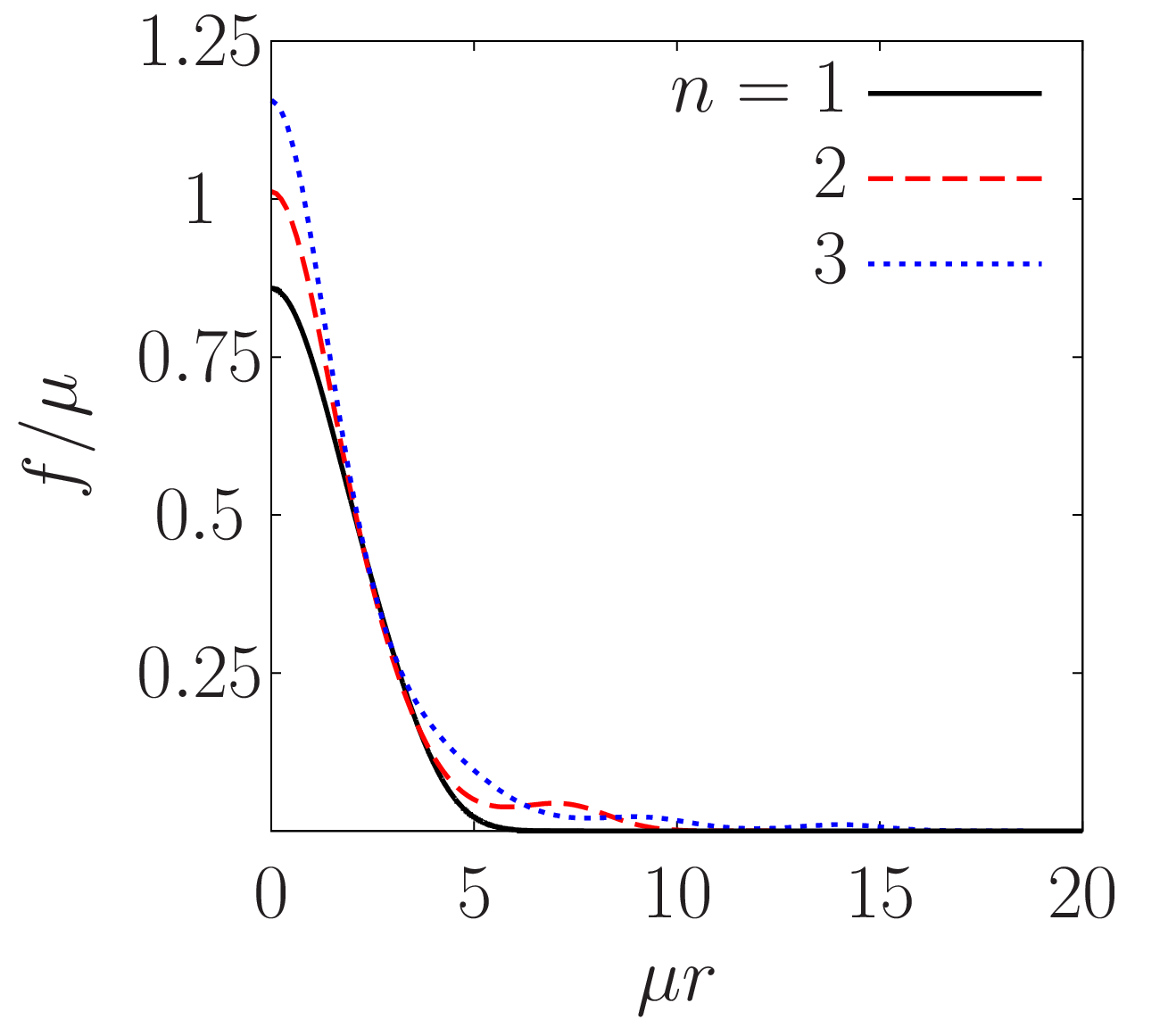}&
\includegraphics[width=0.30\linewidth]{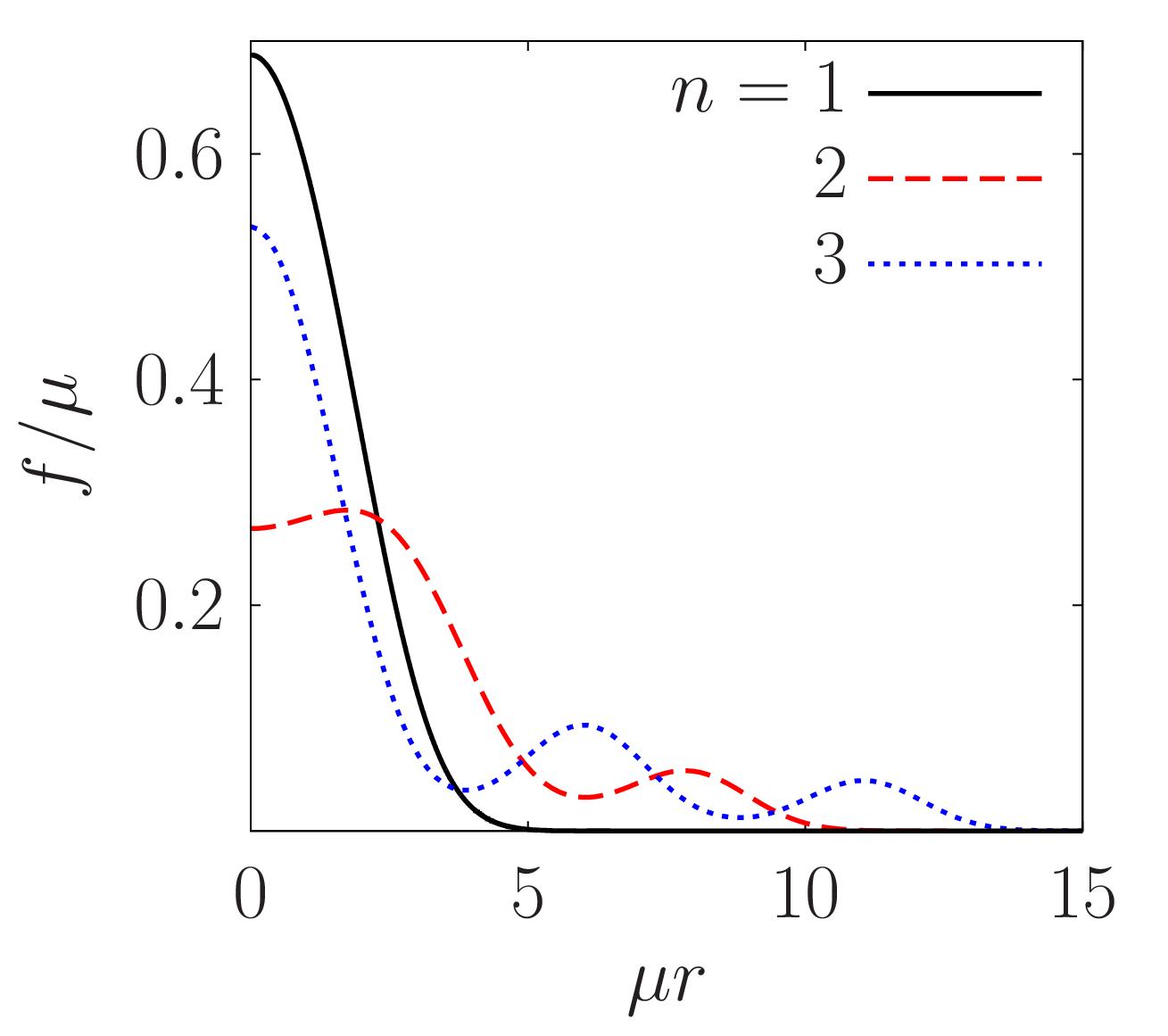}&
\includegraphics[width=0.30\linewidth]{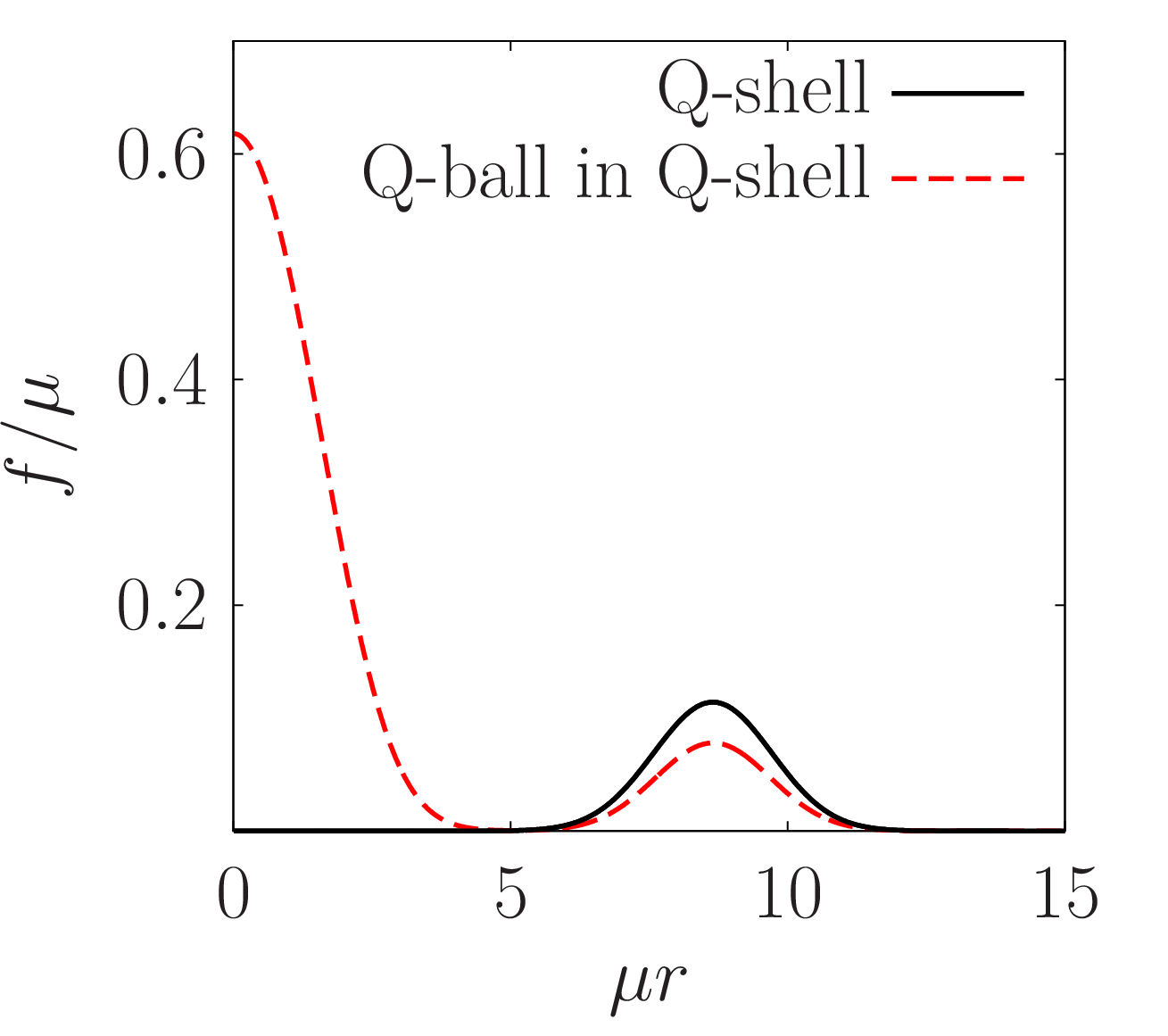}\\
$(a)$&$(b)$&$(c)$\\
\end{tabular}
\caption{Profiles of the gauged Q-ball scalar field $f(r)$ for the solutions marked in Fig.~\ref{figure4} by
asterisks $(a)$, dots $(b)$ and circles $(c)$.}
\label{figure5}
\end{figure}
In Fig.~\ref{figure4}$(a)$ one can find the plane of initial values $f(0)$, $\omega+eA_0(0)$ for the gauged Q-balls in this model (one can see that now even the combination $\omega+eA_0(0)$ does not uniquely characterize a gauged Q-ball). The corresponding $Q(\omega)$ dependence is presented in Fig.~\ref{figure4}$(b)$. The number $n$ labels the number of maxima (global or local) which has the gauged Q-ball scalar field profile $f(r)$. For illustration purposes, we restrict ourselves to solutions with $n \leq 3$. Particular examples of gauged Q-ball solutions, which belong to different families, are presented in Fig.~\ref{figure5}. These solutions are marked by asterisks, dots or circles in Fig.~\ref{figure4}.

It is interesting to note that the number of maxima $n$ does not change along the each family line for solutions located in the right part of the plot in Fig.~\ref{figure4}$(a)$. Meanwhile, solutions located in the left part of the plot in Fig.~\ref{figure4}$(a)$ form the spiral such that $n$ coincides with the number of windings of the spiral.

Another interesting feature of the model with potential \eqref{potential3} is the existence of the so-called Q-shells \cite{Arodz:2008nm} --- solutions with vanishingly small (but positive) values of the scalar field $f(r)$ in the inner part of the solution. There are also solutions which represent gauged Q-balls inside Q-shells (see Fig.~\ref{figure5}$(c)$). In a given model, such solutions may exist if a Q-shell with a rather large radius and a gauged Q-ball are such that the frequency $\omega$ of the gauged Q-ball is almost the same as the value $\omega_{s}+eA_{0}^{s}(0)$ of the Q-shell, representing the continuity condition of the fields at the matching radius. Such solutions are similar to gauged Q-balls with nonmonotonic profile of the scalar field $f(r)$, described above, but with one exception --- for the ``gauged Q-ball inside Q-shell'', the scalar fields of the Q-ball and Q-shell at the matching point (which is a local minimum of the scalar field profile of the ``gauged Q-ball inside Q-shell'') are vanishingly small, which differs from the case of nonmonotonic gauged Q-balls.
\begin{figure}[!ht]
\begin{tabular}{cc}
\includegraphics[width=0.47\linewidth]{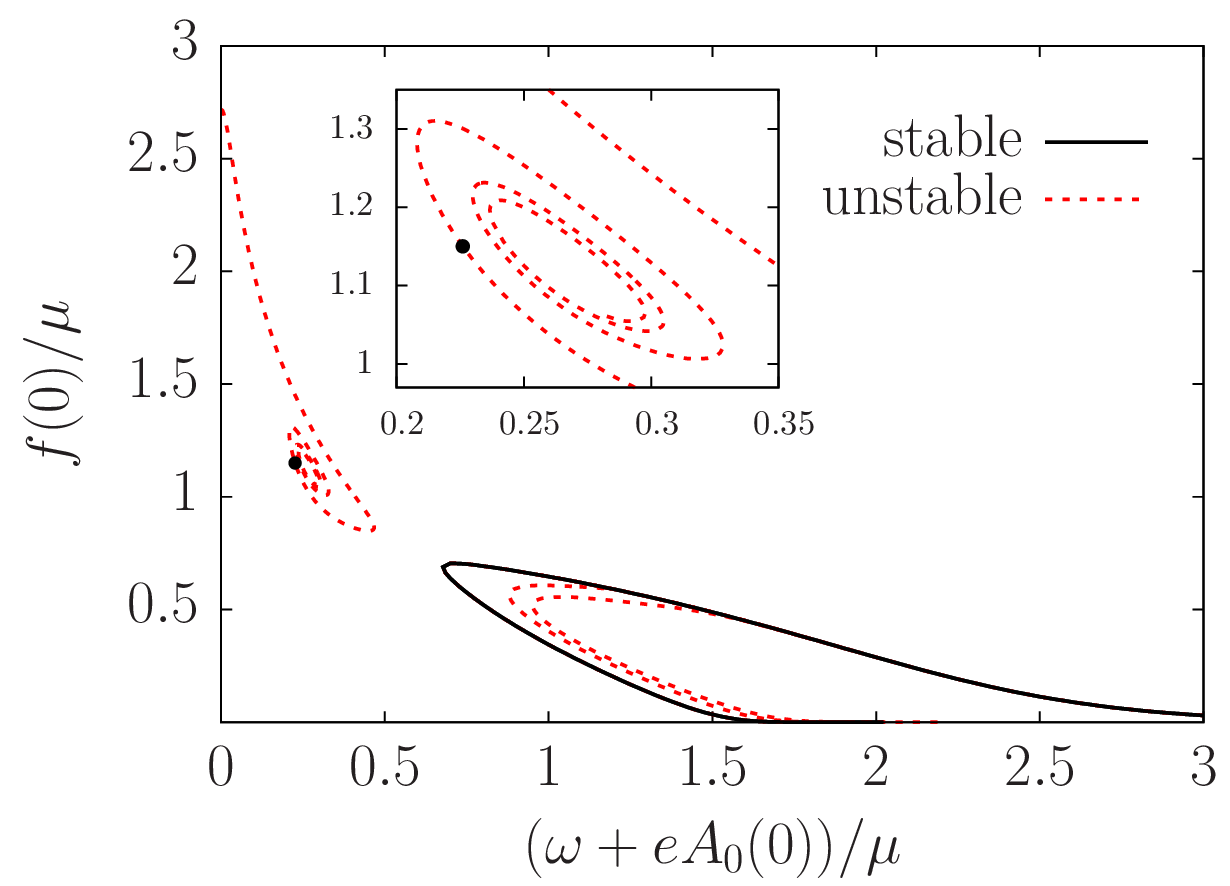}&
\includegraphics[width=0.47\linewidth]{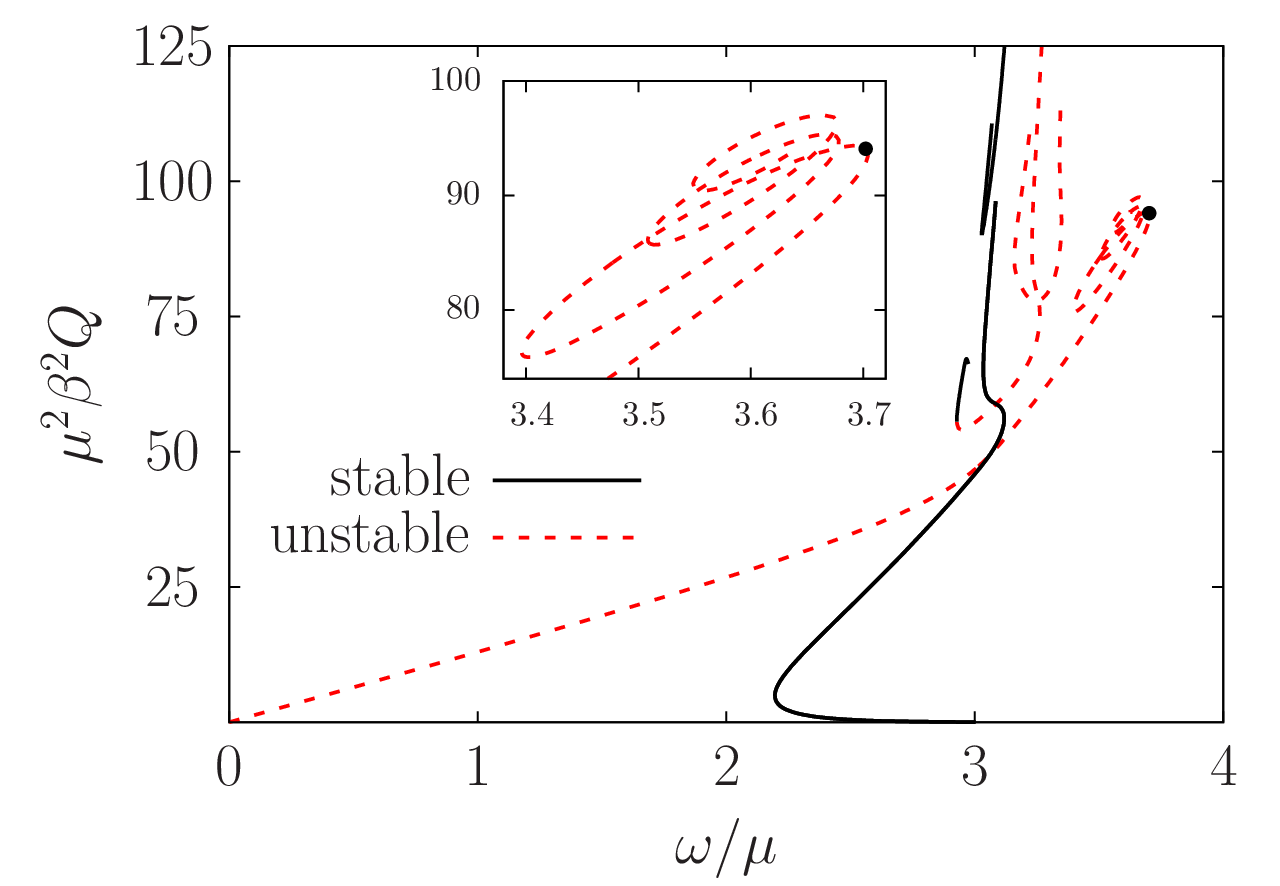}\\
$(a)$&$(b)$\\
\end{tabular}
\caption{Classically stable and classically unstable gauged Q-balls in the model with potential \eqref{potential3} for ${e}/{\beta\mu}=1.1$ on the plane of initial data $(f(0),\,\omega+eA_0(0))$ $(a)$ and on the $Q(\omega)$ dependence $(b)$. Stable solutions are represented by solid lines, unstable solutions are represented by dashed lines. Again, some lines on plot $(a)$, corresponding to visible lines on plot $(b)$ (including the shorter solid lines), appear to be hidden on plot $(a)$ by the curve, corresponding to the longest solid line on plot $(b)$ (see also Fig.~\ref{figure4}).}
\label{figure6}
\end{figure}

\begin{figure}[!ht]
\centerline{\includegraphics[width=0.8\linewidth]{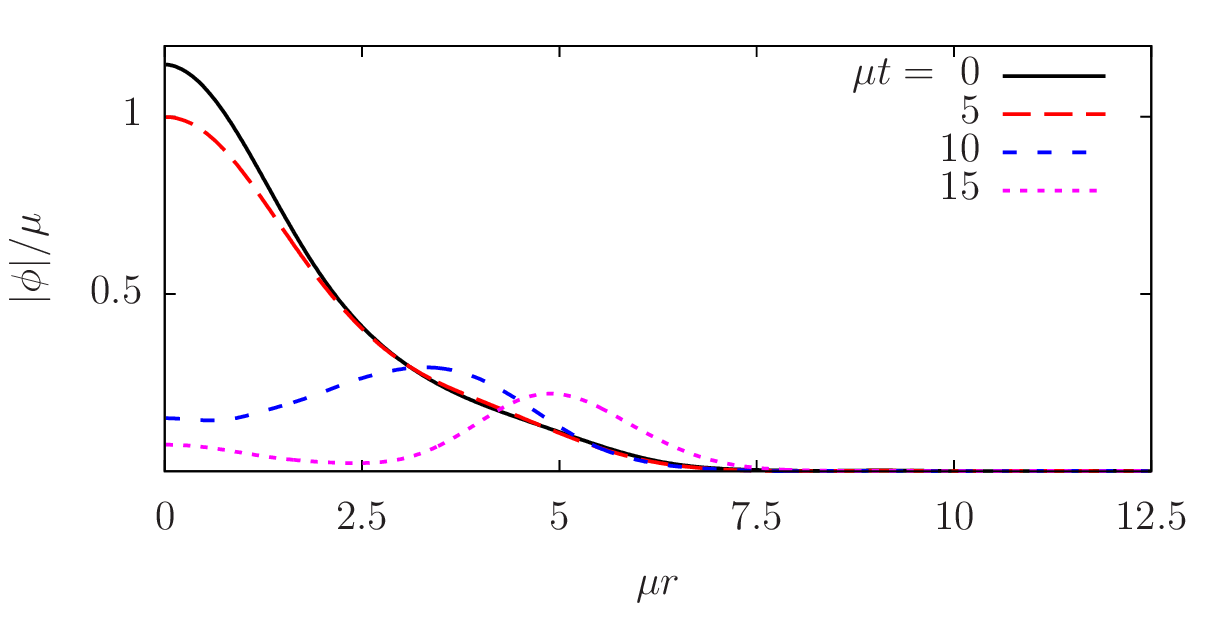}}
\caption{The scalar field profile of the classically unstable gauged Q-ball at different moments of time. The initial solution (at $\mu t=0$) is marked by the dot in Fig.~\ref{figure6}.}
\label{figure7}
\end{figure}

Finally, let us discuss the classical stability of gauged Q-balls in this model for $\frac{e}{\beta\mu}=1.1$. Classically stable and unstable gauged Q-balls (of course, with respect to spherically symmetric perturbations) are marked in Fig.~\ref{figure6} by the solid and dashed lines respectively. One can see from Fig.~\ref{figure6} that there are stable solutions with $\frac{dQ}{d\omega}<0$ and $\frac{dQ}{d\omega}>0$, as well as unstable solutions with $\frac{dQ}{d\omega}<0$ and $\frac{dQ}{d\omega}>0$. At least among the obtained solutions, classically stable gauged Q-balls include only solutions from the family with $n = 1$ and the ``gauged Q-ball inside Q-shell'' solutions. All other solutions are classically unstable. For example, the gauged Q-ball solution, marked by the dot in Fig.~\ref{figure6}, belongs to the family with $n=1$. Evolution in time of this perturbed gauged Q-ball is shown in Fig.~\ref{figure7}, where the scalar field profile is presented at different moments of time. Moreover, $\frac{dQ}{d\omega}<0$ for this Q-ball. This explicit example demonstrates that even in the case of spherically symmetric perturbations, there is no connection between the sign of $\frac{dQ}{d\omega}$ and the classical stability of gauged Q-balls in the general case, which supports the analytical results obtained in Subsection~3.1. Thus, the classical stability criterion for nongauged Q-balls $\frac{dQ}{d\omega}<0$ cannot be applied to gauged Q-balls in the general case.

\section{Conclusion}
In the present paper, we examined the problem of classical stability of $U(1)$ gauged Q-balls. For ordinary (nongauged) Q-balls there exists the well-known classical stability criterion \cite{Friedberg:1976me,LeePang}, stating that Q-balls with $\frac{dQ}{d\omega}<0$ are classically stable if the additional restriction on the number of negative eigenvalues of the corresponding operator (the operator $L_{+}$ in our notations) holds. In principle, the latter condition usually holds for ordinary Q-balls. So, there arises the question of whether this criterion is valid for $U(1)$ gauged Q-balls.

As a preliminary step, we presented the derivation of the classical stability criterion for the one-field and two-fields nongauged Q-balls. Contrary to the original derivation in \cite{Friedberg:1976me,LeePang}, our method is based on the use of only the linearized equations of motion for the perturbations above the gauged Q-ball. In fact, our method is a generalization of the well-known Vakhitov-Kolokolov method \cite{VK,Kolokolov}, which was used to obtain the classical stability criterion for the systems described by the nonlinear Schr\"{o}dinger equation.

This generalized Vakhitov-Kolokolov approach was applied to the case of $U(1)$ gauged Q-balls. However, although all the technical steps of calculations can be performed in the same way as those for ordinary Q-balls (especially, for two-field Q-balls), it is impossible to draw any conclusion about the validity of the classical stability criterion $\frac{dQ}{d\omega}<0$ for the case of $U(1)$ gauged Q-balls. The problem is that, even in the simple case of spherically symmetric perturbations, the structure of the operator $L_{+}$ is such that it is very unlikely that it has only one negative eigenvalue, which is crucial for the validity of the classical stability criterion. In such a case one may conclude that the classical stability criterion $\frac{dQ}{d\omega}<0$, which is valid at least for most of the ordinary (nongauged) Q-balls, cannot be applied to $U(1)$ gauged Q-balls in the general case.

To support the conclusion following from the analytical considerations, we performed numerical simulations of the time evolution of perturbed $U(1)$ gauged Q-balls for the case of spherically symmetric perturbations. The results of simulations demonstrate that there exist classically stable $U(1)$ gauged Q-balls with $\frac{dQ}{d\omega}>0$ and, which is much more important, classically unstable $U(1)$ gauged Q-balls with $\frac{dQ}{d\omega}<0$. At the moment it is not clear whether the $U(1)$ gauged Q-balls, which are classically stable with respect to spherically symmetric perturbations, remain stable in the case of nonspherically symmetric perturbations. In any case, the numerical simulations indeed support the conclusion that the standard classical stability criterion for ordinary (nongauged) Q-balls cannot be applied to $U(1)$ gauged Q-balls in the general case.

\section*{Acknowledgements}
The authors are grateful to E.~Gryzlova, D.~Levkov, E.~Nugaev, and Yu.~Popov for valuable discussions. The work was supported by the Grant 16-12-10494 of the Russian Science Foundation. Numerical calculations were performed on the computational cluster of the theoretical division of INR RAS. We thank G.~Rubtsov for maintaining its stability.

\section*{Appendix~A: Restriction on the form of perturbations for one-field Q-balls}
Let us consider the quantities $\bra{\xi_{2}}L_{2}L_{1}\ket{\xi_{1}}$ and $\bra{\xi_{1}}L_{1}L_{2}\ket{\xi_{2}}$.\footnote{Here, $\bra{f_{1}}\ket{f_{2}}$ is defined as $\bra{f_{1}}\ket{f_{2}}=\int f_{1}^{*}(\vec x)f_{2}(\vec x)d^{d}x$.} From equations \eqref{lineq1} and \eqref{lineq2} we easily get
\begin{eqnarray}\label{twoop1}
\bra{\xi_{2}}L_{2}L_{1}\ket{\xi_{1}}=2\omega\rho\expval{L_{2}}{\xi_{2}}+\rho^{2}\bra{\xi_{2}}L_{2}\ket{\xi_{1}},\\ \label{twoop2}
\bra{\xi_{1}}L_{1}L_{2}\ket{\xi_{2}}=2\omega\rho\expval{L_{1}}{\xi_{1}}+\rho^{2}\bra{\xi_{1}}L_{1}\ket{\xi_{2}}.
\end{eqnarray}
Again, using equations \eqref{lineq1} and \eqref{lineq2} we get
\begin{eqnarray}\label{firsteq}
&&\expval{L_{2}}{\xi_{2}}=2\omega\rho\bra{\xi_{2}}\ket{\xi_{1}}+\rho^{2}\bra{\xi_{2}}\ket{\xi_{2}},\\
&&\bra{\xi_{2}}L_{2}\ket{\xi_{1}}=2\omega\rho^{*}\bra{\xi_{1}}\ket{\xi_{1}}+{\rho^{*}}^{2}\bra{\xi_{2}}\ket{\xi_{1}},\\ \label{thirdeq}
&&\expval{L_{1}}{\xi_{1}}=2\omega\rho^{*}\bra{\xi_{2}}\ket{\xi_{1}}+{\rho^{*}}^{2}\bra{\xi_{1}}\ket{\xi_{1}},\\ \label{fourttheq}
&&\bra{\xi_{2}}L_{1}\ket{\xi_{1}}=2\omega\rho\bra{\xi_{2}}\ket{\xi_{2}}+\rho^{2}\bra{\xi_{2}}\ket{\xi_{1}}.
\end{eqnarray}
Equation \eqref{thirdeq} was obtained using the fact that $\expval{L_{1}}{\xi_{1}}^{\dagger}=\expval{L_{1}}{\xi_{1}}$. Using the obvious relation $\bra{\xi_{1}}L_{1}L_{2}\ket{\xi_{2}}^{\dagger}=\bra{\xi_{2}}L_{2}L_{1}\ket{\xi_{1}}$, from \eqref{twoop1} and \eqref{twoop2} we obtain
\begin{equation}\label{equalityL1L2}
2\omega\rho\expval{L_{2}}{\xi_{2}}+\rho^{2}\bra{\xi_{2}}L_{2}\ket{\xi_{1}}=2\omega\rho^{*}\expval{L_{1}}{\xi_{1}}+{\rho^{*}}^{2}\bra{\xi_{2}}L_{1}\ket{\xi_{1}}.
\end{equation}
Substituting equations \eqref{firsteq}--\eqref{fourttheq} into \eqref{equalityL1L2} and multiplying the result by $\rho\neq 0$, we arrive at
\begin{equation}\label{pertconstraint}
\left(\rho^{2}-{\rho^{*}}^2\right)\Bigl(2\omega\bra{\xi_{2}}\ket{\xi_{1}}+\rho\bra{\xi_{2}}\ket{\xi_{2}}+\rho^{*}\bra{\xi_{1}}\ket{\xi_{1}}\Bigr)=0.
\end{equation}

Equation \eqref{pertconstraint} is fulfilled if one of the following simpler equations fulfills
\begin{eqnarray}\label{pertconstraint1}
&&\rho^{2}={\rho^{*}}^2,\\ \label{pertconstraint2}
&&2\omega\bra{\xi_{2}}\ket{\xi_{1}}+\rho\bra{\xi_{2}}\ket{\xi_{2}}+\rho^{*}\bra{\xi_{1}}\ket{\xi_{1}}=0.
\end{eqnarray}
Let us consider the second equation. Equation \eqref{firsteq} together with \eqref{pertconstraint2} results in
\begin{equation}\label{constraintlin}
\expval{L_{2}}{\xi_{2}}+\rho^{*}\rho\bra{\xi_{1}}\ket{\xi_{1}}=0.
\end{equation}
According to equation \eqref{eqqball}
\begin{equation}
L_{2}f=0.
\end{equation}
By the definition, the function $f(r)>0$ for any $r$, i.e., it has no nodes. That means that $f$ is the eigenfunction of the lowest eigenstate (in the case $d=3$ it corresponds to the $1s$ level in the spherically symmetric quantum mechanical potential $U(r)-\omega^{2}+\gamma^{2}$), and there are no negative eigenvalues of the operator $L_{2}$. Thus, we get $\expval{L_{2}}{\xi_{2}}\ge 0$ for any $\xi_{2}$. Since $\bra{\xi_{1}}\ket{\xi_{1}}\ge 0$ and $\rho^{*}\rho\ge 0$, except the trivial solution $\xi_{1}\equiv 0$, $\xi_{2}\equiv 0$, equation \eqref{constraintlin} has the following solutions:
\begin{eqnarray}
&&\xi_{2}\sim f(r),\quad \rho=0\\ \nonumber &&\textrm{or}\\
&&\xi_{2}\sim f(r),\quad \xi_{1}\equiv 0.
\end{eqnarray}
The first solution obviously does not indicate any instability, whereas the second solution together with equation \eqref{lineq2} also leads to $\rho=0$.

Finally, the only case, which can describe possible unstable modes (i.e., those with $\textrm{Im}\,\rho\neq 0$), is defined by equation \eqref{pertconstraint1}.

\section*{Appendix~B: Restriction on the form of perturbations for two-field Q-balls}
Let us consider the following general form of perturbations above the two-field Q-ball:
\begin{eqnarray}
&&\phi(t,\vec x)=e^{i\omega t}f(r)+e^{i\omega t}\left(a(\vec x)e^{i\rho t}+b(\vec x)e^{-i\rho^{*} t}\right),\\
&&\chi(t,\vec x)=g(r)+c(\vec x)e^{i\rho t}+c^{*}(\vec x)e^{-i\rho^{*} t}.
\end{eqnarray}
The corresponding linearized equations of motion can be represented as
\begin{eqnarray}\label{eq1pert2field}
&&L_{1}\Xi_{1}=\rho^{2}\Xi_{1}+2\omega\rho\,\Xi_{2}-\rho^{2}\Xi_{c},\\ \label{eq2pert2field}
&&L_{2}\Xi_{2}=\rho^{2}\Xi_{2}+2\omega\rho\,\Xi_{1}-4\omega\rho\,\Xi_{c},
\end{eqnarray}
where
\begin{equation}
\Xi_{1}=\begin{pmatrix}
a+b^{*} \\ 2c
\end{pmatrix},\qquad \Xi_{2}=\begin{pmatrix}
a-b^{*} \\ 0
\end{pmatrix},\qquad \Xi_{c}=\begin{pmatrix}
0 \\ c
\end{pmatrix}
\end{equation}
and
\begin{equation}
\qquad L_{1}=
\begin{pmatrix}
L_{u} & Y \\
Y & L_{\varphi}
\end{pmatrix},\qquad L_{2}=\begin{pmatrix}
-\Delta+U-\omega^{2} & 0 \\
0 & 0
\end{pmatrix}
\end{equation}
with $L_{u}$, $L_{\varphi}$, $Y$, $U$ defined by \eqref{Lu}, \eqref{Lvarphi}, \eqref{U2field}, and \eqref{W2field}. Again, as in the one-field case, from equations \eqref{eq1pert2field} and \eqref{eq2pert2field} we can get\footnote{The terms of the form $\bra{F_{1}}O\ket{F_{2}}$ are defined as $\bra{F_{1}}O\ket{F_{2}}=\int (F_{1}^{\dagger}O F_{2})d^{d}x$}
\begin{eqnarray}\label{twoop12field}
&&\bra{\Xi_{2}}L_{2}L_{1}\ket{\Xi_{1}}=2\omega\rho\expval{L_{2}}{\Xi_{2}}+\rho^{2}\bra{\Xi_{2}}L_{2}\ket{\Xi_{1}},\\ \label{twoop22field}
&&\bra{\Xi_{1}}L_{1}L_{2}\ket{\Xi_{2}}=2\omega\rho\expval{L_{1}}{\Xi_{1}}+\rho^{2}\bra{\Xi_{1}}L_{1}\ket{\Xi_{2}}-4\omega\rho\bra{\Xi_{1}}L_{1}\ket{\Xi_{c}}.
\end{eqnarray}
Using equations \eqref{eq1pert2field} and \eqref{eq2pert2field}, we obtain
\begin{eqnarray}\label{zeroeq2fields}
&&\bra{\Xi_{1}}L_{1}\ket{\Xi_{c}}={\rho^{*}}^{2}\bra{c}\ket{c},\\ \label{firsteq2fields}
&&\expval{L_{2}}{\Xi_{2}}=2\omega\rho\bra{\Xi_{2}}\ket{\Xi_{1}}+\rho^{2}\bra{\Xi_{2}}\ket{\Xi_{2}},\\
&&\bra{\Xi_{2}}L_{2}\ket{\Xi_{1}}=2\omega\rho^{*}\bra{\Xi_{1}}\ket{\Xi_{1}}+{\rho^{*}}^{2}\bra{\Xi_{2}}\ket{\Xi_{1}}-8\omega\rho^{*}\bra{c}\ket{c},\\ \label{thirdeq2fields}
&&\expval{L_{1}}{\Xi_{1}}=2\omega\rho^{*}\bra{\Xi_{2}}\ket{\Xi_{1}}+{\rho^{*}}^{2}\bra{\Xi_{1}}\ket{\Xi_{1}}-2{\rho^{*}}^{2}\bra{c}\ket{c},\\ \label{fourttheq2fields}
&&\bra{\Xi_{2}}L_{1}\ket{\Xi_{1}}=2\omega\rho\bra{\Xi_{2}}\ket{\Xi_{2}}+\rho^{2}\bra{\Xi_{2}}\ket{\Xi_{1}}.
\end{eqnarray}
Equation \eqref{thirdeq2fields} was obtained using the fact that $\expval{L_{1}}{\Xi_{1}}^{\dagger}=\expval{L_{1}}{\Xi_{1}}$. Using the obvious relation $\bra{\Xi_{1}}L_{1}L_{2}\ket{\Xi_{2}}^{\dagger}=\bra{\Xi_{2}}L_{2}L_{1}\ket{\Xi_{1}}$, from \eqref{twoop12field} and \eqref{twoop22field} we obtain
\begin{equation}\label{equalityL1L22fields}
2\omega\rho\expval{L_{2}}{\Xi_{2}}+\rho^{2}\bra{\Xi_{2}}L_{2}\ket{\Xi_{1}}=2\omega\rho^{*}\expval{L_{1}}{\Xi_{1}}+{\rho^{*}}^{2}\bra{\Xi_{2}}L_{1}\ket{\Xi_{1}}
-4\omega{\rho^{*}}\rho^{2}\bra{c}\ket{c}.
\end{equation}
Substituting equations \eqref{zeroeq2fields}--\eqref{fourttheq2fields} into \eqref{equalityL1L22fields}, we arrive at
\begin{equation}\label{pertconstraint2fields}
\left(\rho^{2}-{\rho^{*}}^2\right)\Bigl(2\omega\bra{\Xi_{2}}\ket{\Xi_{1}}+\rho\bra{\Xi_{2}}\ket{\Xi_{2}}+\rho^{*}\bra{\Xi_{1}}\ket{\Xi_{1}}
-2\rho^{*}\bra{c}\ket{c}\Bigr)=0.
\end{equation}
Multiplying the latter equation by $\rho\neq 0$ and using equation \eqref{firsteq2fields}, we arrive at
\begin{equation}
\left(\rho^{2}-{\rho^{*}}^2\right)\Bigl(\expval{L_{2}}{\Xi_{2}}+\rho^{*}\rho\bra{\Xi_{1}}\ket{\Xi_{1}}
-2\rho^{*}\rho\bra{c}\ket{c}\Bigr)=0.
\end{equation}
Using the definition of $\Xi_{1}$ and $\Xi_{2}$, finally we get
\begin{equation}
\left(\rho^{2}-{\rho^{*}}^2\right)\Bigl(\expval{-\Delta+U-\omega^{2}}{\xi_{2}}+\rho^{*}\rho\bra{\xi_{1}}\ket{\xi_{1}}+2\rho^{*}\rho\bra{c}\ket{c}\Bigr)=0,
\end{equation}
where
\begin{equation}
\xi_{1}=a+b^{*},\qquad \xi_{2}=a-b^{*}.
\end{equation}
Since, analogously to the one-field case, $\expval{-\Delta+U-\omega^{2}}{\xi_{2}}\ge 0$ for any $\xi_{2}$, nontrivial solutions with $\rho\neq 0$ should satisfy the condition
\begin{equation}
\rho^{2}={\rho^{*}}^2,
\end{equation}
leading to $\rho=\gamma$ or $\rho=i\gamma$, where $\gamma$ is a real constant. Thus, unstable modes can have only the form \eqref{pertdef2fields} and \eqref{pertchidef2fields}.

\section*{Appendix~C: Numerical method}
Here we present a more detailed description of the numerical method used to solve the equations of motion \eqref{numeq1} and \eqref{numeq2}.
First, we introduce the new variable
\begin{equation}
\label{newr}
{\tilde r} = \frac{r}{r+r_0},
\end{equation}
which maps the infinite space on the unit size sphere. Here $r_0$ is a constant parameter. With this coordinate, the boundary conditions \eqref{bc1} and \eqref{bc2} appear to be imposed at the points ${\tilde r} = 0$ and ${\tilde r} = 1$. In our simulations, we define the fields on the uniform grid with $N = 2000$ points which cover this interval. Note that in terms of the physical distance the grid is not uniform. Half of the grid points cover the range $r \in [0,r_0]$, where the grid is approximately uniform with the spacing varying from $\triangle r = r_0 \,\triangle {\tilde r}$ to $\triangle r = 4 r_0 \,\triangle {\tilde r}$. The remaining points cover the interval $r \in [r_0,\infty)$ with the spacing $\triangle r \simeq r^2\triangle {\tilde r}/r_0$ for $r\gg r_{0}$. In our simulations, the parameter $r_0$ is chosen to be of the order of twice the Q-ball radius, so we have approximately one and the same spacing $\triangle r$ in the region of the Q-ball core. We use the second-order central finite difference formulas to discretize the derivatives with respect to time $\tilde t$ (here $\tilde t$ is the dimensionless variable corresponding to the physical time $t$) and radius $\tilde{r}$ in the equations of motion for the fields.

Outgoing waves leaving the interaction region are removed with the help of the Kreiss-Oliger filter \cite{KreissOliger}. After each time step $\triangle\tilde t$ we apply the following operator
\begin{equation}
\phi \to \left(1 - \frac{\epsilon\,\triangle\tilde t\, (\triangle {\tilde r})^4}{2^4} \partial_{\tilde
  r}^{4} \right) \phi,
\end{equation}
to the field $\phi$, where $\epsilon<1$ is the constant introduced to maintain the stability of this procedure. For solutions which are smooth enough, this procedure modifies the scalar field by the fourth-order term, which is smaller than the error of the numerical scheme in the region of the core of the gauged Q-ball.

To demonstrate how this operator works, let us consider its action (with the derivative $\partial_{\tilde r}^{4}$ discretized by the second-order central finite difference formula) on the Fourier modes of the field $\phi$. We get
\begin{equation}
\label{KO-k}
\phi_{\tilde k} \to \left(1 - \epsilon\, \triangle\tilde t \sin^{4}\left(\frac{{\tilde k}\triangle{\tilde r}}{2}\right)\right) \phi_{\tilde k}\;.
\end{equation}
One can see that the low frequency modes with ${\tilde k} \ll 1/ \triangle {\tilde r}$ remain practically intact; their amplitudes appear to be modified by the factor $(1 - \epsilon\, \triangle\tilde t\, ({\tilde k} \triangle {\tilde r}/2)^4)$, which is very close to unity. This is the case of the modes located in the central region $r \in [0,r_0]$ in the physical space, where the grid resolution is high enough. As for the modes ${\tilde k} \sim 1/ \triangle {\tilde r}$, the factor in \eqref{KO-k} is less than unity in this case and the amplitudes of these modes vanish after $\sim 1/{\triangle\tilde t}$ time steps. Roughly speaking, such modes correspond to the scalar waves which propagate in the physical space out from the Q-ball core and appear in the region of the physical space where the grid is sparse. This method is similar to the one used in \cite{Honda:2001xg}.

As was mentioned above, the derivatives in the equations of motion are discretized by the second-order central finite difference formulas. The resulting system of the nonlinear equations of motion is solved by the variant of the iterated Crank-Nicolson method used in \cite{Teukolsky:1999rm}. Namely, for the time step $\tilde t_{n}$, $\tilde t_{n}+\triangle\tilde t$ we make iterations such that in each iteration the gauge field $A_{0}$ and the terms $\partial_{t}A_{0}$,  $\frac{dV}{d(\phi^{*}\phi)}$ in equation \eqref{numeq1} and the terms $\left(\phi^{*}\partial_{t}\phi-\phi\partial_{t}\phi^{*}\right)$, $\phi\phi^{*}$ in equation \eqref{numeq2} (but not the terms with spatial derivatives in both equations) are calculated using the data obtained in the previous iteration according to the scheme presented in \cite{Teukolsky:1999rm} (for the first iteration, we use the data from the previous time step for the time derivatives $\partial_{t}A_{0}$ in \eqref{numeq1} and $\partial_{t}\phi$, $\partial_{t}\phi^{*}$ in \eqref{numeq2}). Then, these terms are considered as the coefficients depending only on the spatial coordinate $\tilde r$, so formally we solve the linear equations of motion (homogeneous for the scalar field $\phi$ and inhomogeneous for the gauge field $A_{0}$) using the implicit Crank-Nicolson scheme. After each iteration, we get more accurate results for the fields and more accurate values of the corresponding terms in the equations of motion, which are used for the next iteration. We find that three iterations are enough to achieve the accuracy of the solution which is better than the accuracy of the discretization. Since the scheme is implicit, it is stable with respect to von Neumann stability analysis for more than two iterations \cite{Teukolsky:1999rm}.

\end{document}